# LLM for SoC Security: A Paradigm Shift

Dipayan Saha, Shams Tarek, Katayoon Yahyaei, Sujan Kumar Saha, Jingbo Zhou,
Mark Tehranipoor, and Farimah Farahmandi

*Department of Electrical and Computer Engineering, University of Florida, Gainesville, FL, USA.*

Email: {dsaha, shams.tarek, ka.yahyaei, sujansaha, jingbozhou}@ufl.edu, {tehranipoor, farimah}@ece.ufl.edu

*Abstract*—As the ubiquity and complexity of system-on-chip (SoC) designs increase across electronic devices, the task of incorporating security into an SoC design flow poses significant challenges. Existing security solutions are inadequate to provide effective verification of modern SoC designs due to their limitations in scalability, comprehensiveness, and adaptability. On the other hand, Large Language Models (LLMs) are celebrated for their remarkable success in natural language understanding, advanced reasoning, and program synthesis tasks. Recognizing an opportunity, our research delves into leveraging the emergent capabilities of Generative Pre-trained Transformers (GPTs) to address the existing gaps in SoC security, aiming for a more efficient, scalable, and adaptable methodology. By integrating LLMs into the SoC security verification paradigm, we open a new frontier of possibilities and challenges to ensure the security of increasingly complex SoCs. This paper offers an in-depth analysis of existing works, showcases practical case studies, demonstrates comprehensive experiments, and provides useful promoting guidelines. We also present the achievements, prospects, and challenges of employing LLM in different SoC security verification tasks.

*Index Terms*—Large Language Model, ChatGPT, GPT-4, Hardware Security, SoC Security Verification, RTL

## I. INTRODUCTION

The recent rise of Large Language Models (LLMs) has profoundly impacted the field of Natural Language Processing (NLP), ushering in a new era of capabilities and applications. As the size and complexity of these models increase, they consistently improve in performance and efficiency on numerous NLP tasks that span Natural Language Generation (NLG) [1], Natural Language Understanding (NLU) [2] and information retrieval [3]. Specifically, their mastery is evident in fields such as text generation [4], summarization [5, 6], machine translation [7], paraphrasing [8], classification [9], sentiment analysis [10], and question answering [11], to name a few. Beyond their efficacy in such linguistic tasks, LLMs are increasingly showcasing incredible aptitude in complex reasoning tasks. This encompasses arithmetic reasoning [12], commonsense, symbolic, and logical deliberations [13], analogical reasoning [14], and even multimodal reasoning [15]. Such emergent abilities [16], more pronounced in larger models such as GPT-3 [17], GPT-4 [18], PaLM [19], etc., provide a captivating insight into the unforeseen potential of scaled-up language models. Because of zero-shot and few-shot learning capabilities, these pre-trained models (PTMs) are being applied in a wide range of applications: healthcare [20], legal professions [21–23], creative works [24], and robotics [15, 25]. The remarkable success of these PTMs has catalyzed the development of fine-tuned domain-specific LLMs such as Med-PaLM [26], Med-PaLM 2 [27], PaLM-E [25], BloombergGPT [28], AugGPT [29], LayoutGPT [30], BioBERT [31], SciBERT [32], ClimateBERT [33], etc.

As an example, software programming is witnessing a transformative shift as researchers increasingly incorporate LLMs for diversified coding tasks. Code LLM models [34–43], with their deep understanding of code syntax, semantics, and the intricacies of various programming languages, are being deployed to assist in code generation, completion, translation, explanation, and documentation. Such capabilities streamline the coding process, reduce the margin for human error, and increase overall efficiency. For example, GitHub Copilot [44], based on Codex [35], offers contextual suggestions for multiple programming languages, bridging the gap between human intuition and machine efficiency in software development. Following the remarkable successes in coding assistance, LLMs are expanding their horizons to address the pressing challenges of software security. Recent studies [45–56] demonstrate the ability of LLMs to identify and fix software bugs based on natural language description.

The widespread presence of system-on-chip (SoC) in modern computing systems emphasizes its critical importance. SoCs are now integrated into diverse devices, including smartphones, tablets, IoT devices, and autonomous vehicles, showcasing their significance in the technology landscape. With such an increase in their use, security has become an increasing concern as SoCs collect, analyze, and store users' personal information. Multiple intellectual property (IP) cores with unique functionality and security challenges come together to make an SoC. The extensive functionality coupled with intricate interactions among the IPs, leaves SoCs susceptible to a plethora of security vulnerabilities. From these vulnerabilities, adversaries can exploit information leakage [57, 58], side-channel leakage [59–61], access control violations [62], etc. The situation is further complicated when considering third-party IPs, which are notably prone to issues like hardware Trojans [63]. These issues highlight the importance of thorough security verification in system design. This rigorous and time-intensive process is at odds with the escalating demand for producing billions of computing devices and the corresponding pressure to reduce time-to-market [64]. The tension between these opposing factors makes effective functionality and security verification increasingly difficult, potentially leading to costly spin-offs if issues are discovered post-production. Unfortunately, the existing SoC security solutions [62, 65–69] are not scalable for handling the increasing complexity and di-versity of modern hardware designs, adaptable to new designs and rapidly evolving threat landscape, and comprehensive in addressing hardware vulnerabilities[70, 71].

Given the complexity and diversity of SoC security issues and the proven prowess of LLMs in coding, NLU, and advanced reasoning activities, the idea of integrating LLMs into the SoC security paradigm appears promising. Such an interaction holds the potential to not only address the existing challenges in SoC security but also to pioneer innovative solutions for the future with the help of the emergent abilities of LLMs. Moreover, the hardware security community has recently begun to explore the potential of LLMs for SoC security [72–78]. These endeavors, which target specific individual challenges within hardware security, highlight the promise of LLMs in this domain. Nevertheless, the amount of existing research in this domain is inadequate and the real potential of LLMs in different SoC security tasks is untapped. This is the first of its kind work that addresses this research gap, thoroughly investigating the potential of LLMs in SoC security verification.

Figure 1 presents a comprehensive illustration of the potential applications of LLM in SoC security, as addressed in this work. The potential of LLM, combined with the proper selection of learning paradigm, the finesse of prompt engineering, and the rigor of fidelity checks, holds the promise of redefining security tasks across domains. Within this context, we explore the following four different security tasks:

1) **Vulnerability Insertion**: We show how adeptly LLM can introduce potential vulnerability and weakness into RTL design following natural language description through the guidance of a well-crafted prompt.

2) **Security Assessment**: Through security assessment, we harness the prowess of LLMs to critically evaluate the security landscape of hardware designs to identify vulnerabilities, weaknesses, and threats through LLM. We also examine the ability of LLM to pinpoint simple coding issues that can turn into security bugs.

3) **Security Verification**: In this scenario, we use LLM to verify if the design meets specific security rules or policies. Furthermore, we check the proficiency of LLM in calculating security metrics, understanding security properties, and generating functional testbenches to identify weaknesses.

4) **Countermeasure Development**: In this scenario, we analyze how effectively LLM can mitigate the existing vulnerabilities embedded in the design.

In each of the outlined scenarios, we provide a comprehensive demonstration of executing the tasks using LLMs, with an emphasis on the strategic use of prompt engineering. Through these practical case studies, we establish several strong prompt guidelines specifically applicable to each of the security tasks. In addition, by conducting extensive evaluations, we investigate the proficiency of specific LLMs—particularly GPT-3.5 and GPT-4—in undertaking these four critical security tasks. Our exhaustive discussions and empirical findings not only note the successes of LLMs in the SoC security landscape thus far, but also pinpoint the prospects and prevailing challenges of employing LLMs in SoC security.

To both the scholarly community and industry professionals, this research work acts as a foundational guide, laying the groundwork for the integration of LLMs in the SoC security landscape. Our exhaustive survey of existing LLMs and related security works not only informs readers of the current state-of-the-art but also explains the evolutionary trajectory of these models, aiding in understanding their capabilities and potential applications in the abovementioned tasks. Our observations and insights through discussions, case studies, and experimentation will help the reader understand the advantages and challenges of employing LLM in SoC security. The comprehensive nature of our investigation opens doors to further exploration, encouraging the community to dive deeper, innovate, and push the boundaries of what is possible, ultimately strengthening the foundations and advancements in hardware security.

The contributions of this work are listed below:

- This is the first of its kind that thoroughly investigates the potential of LLMs in different SoC security-related tasks.
- We provide a comprehensive survey of existing LLMs and related work.
- We first formulate key research questions regarding the prospects of LLM in SoC security and later, systematically analyze them in-depth through practical case studies and large-scale investigation.
- We identify several specific prompt guidelines for using LLM effectively in SoC security-related tasks.
- We identify potential challenges of using LLM in SoC security and also note prospects for further research.

In the remainder of this paper, Section II narrates the preliminaries on SoC security. Later, Section III describes the interaction between LLM and SoC security by providing a comprehensive survey of existing LLMs and different related aspects. Eight research questions addressed in this work are described in Section IV, followed by seven case studies in Section V. Afterward, large-scale investigations on different capabilities of LLM in SoC security-related tasks are discussed in Section VI. Finally, Section VII concludes the paper.

## II. SoC Security

Contemporary SoCs are progressively reaching higher levels of advancement and intricacy. However, this increased complexity also opens up concealed vulnerabilities that attackers can take advantage of. Any bug inside the SoC should be caught and fixed at the earliest stage of the design flow; otherwise, it will be 10 times more expensive than the previous stage. There are rules and guidelines to address the problem of functional bugs in the design-time [79]. The RTL design is the first stage of the design flow in modern ICs. Therefore, being able to detect and mitigate all the possible vulnerabilities at this stage will significantly decrease the security verification time, effort, and cost.

*1) Security Vulnerabilities:* Hardware security vulnerabilities are the points in design that an attacker could exploit to gain access to a resource with a security-critical value. A vulnerability can appear in the design in the following ways:

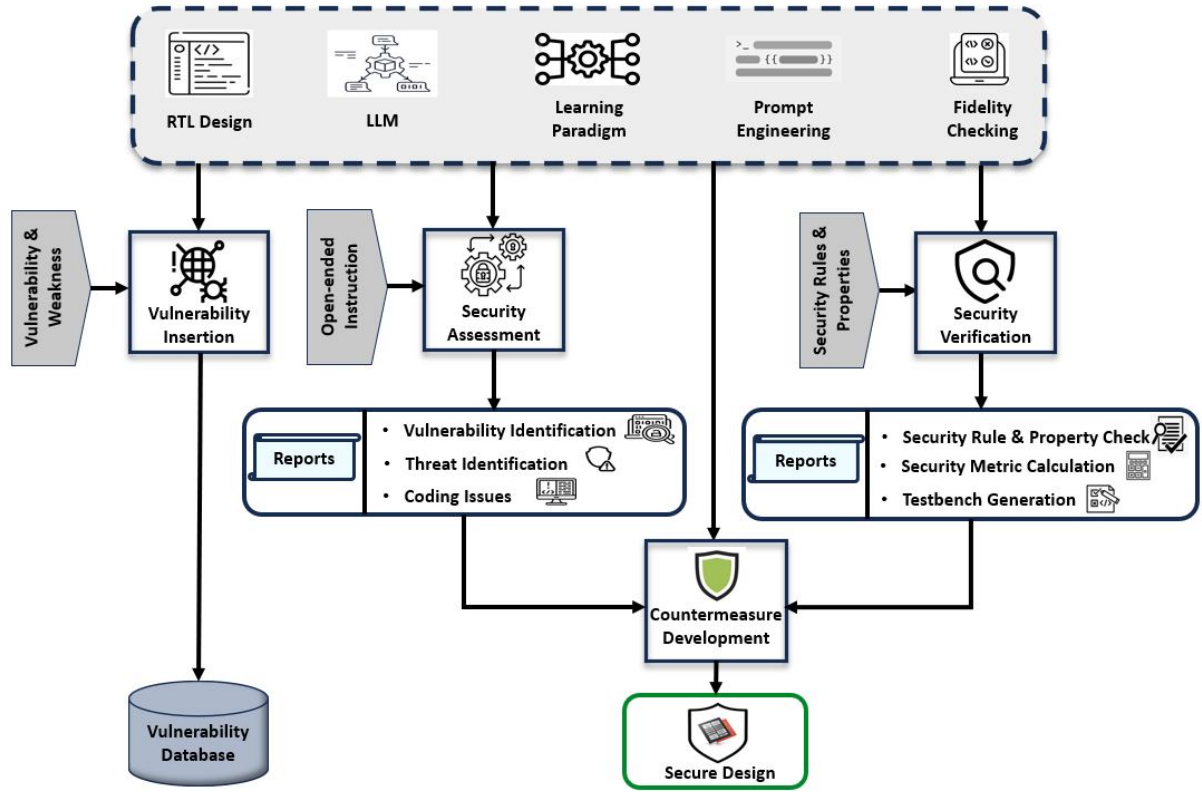

Fig. 1. Potential applications of LLM in SoC security.

- Designers' mistsakes: Designers may not fully understand the implications of their choices or make mistakes in the design phase. As a result, they can inadvertently create security vulnerabilities.
- Vulnerabilities through automation tools: As designs move from abstract models to specific implementations, computer-aided design (CAD) tools may unintentionally introduce vulnerabilities during synthesis and optimization [59].
- Malicious Modifications: Rogue designers or third-party IP providers can intentionally introduce vulnerabilities that create backdoors that result in unauthorized access, alterations, and control of the system [59].
- Test and Debug Infrastructure: Sometimes enhancements in control and observation capabilities for testing and debugging can also provide vectors for attackers to violate confidentiality and integrity in the post-silicon stages [58].

Figure 2 shows the possible locations of the hardware vulnerabilities inside a typical SoC. Many of these hardware vulnerabilities could be prevented or mitigated during design time with a set of guidelines for general hardware description language (HDL) code structures and design practices. The previously discovered vulnerabilities can be utilized for studying the root cause of them in the design and eventually their mitigation. The most relevant work on hardware vulnerability is Common Weakness Enumeration (CWE). It is a community-developed list of hardware and software weaknesses founded by MITRE [80]. CWE offers an initial approach to catego-rizing various hardware vulnerabilities through a set of three fundamental questions: the underlying reason for including the vulnerability in the design, the timing of its inclusion, and its potential location. The database is continuously maintained through regular assessments of introduced vulnerabilities and is further organized into distinct categories for the user. For example, CWE-1260 specifically addresses issues related to improper memory overlap among protected memory regions. If an application operating at a lower privilege level is intentionally programmed to overlap with an application at a higher privilege level, it can result in a privilege escalation problem, potentially leading to a security breach. While CWE vulnerabilities offer a fundamental understanding of hardware vulnerabilities, they lack benchmarks to validate vulnerable designs.

Another comprehensive resource is the Trust-Hub property database [81]. Additionally, hackathons like Hack@Dac [82] and the HOST Microelectronics Challenge [83] serve as valuable resources for identifying emerging vulnerabilities in SoC designs. However, many of these documented vulnerabilities are either not open-source or lack adequate documentation. The work presented in [84] provides a comprehensive database of SoC vulnerability benchmarks, facilitating the verification of different designs using various verification techniques on a common platform. Nevertheless, there is still room for significant improvement, as new vulnerabilities are continually discovered and can pose significant security risks to hardware designs.

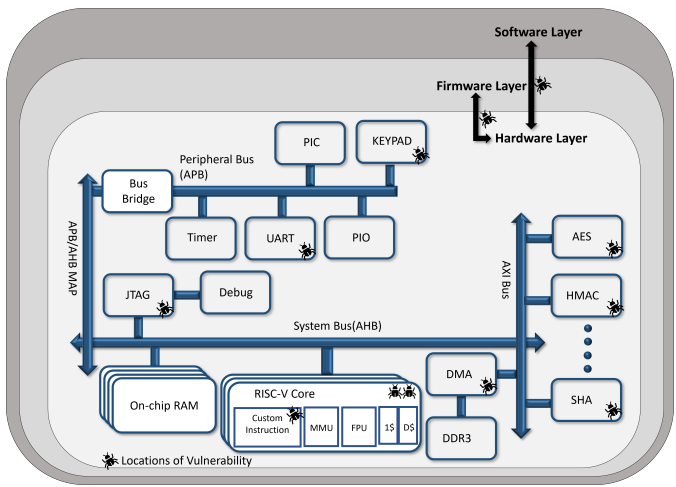

Fig. 2. A typical SoC with probable locations of hardware bugs

*2) Threat Modeling:* Threat modeling in hardware security involves identifying potential adversaries such as hackers, insider threats, competitors, or malicious entities, and the assets they might target, like IP or sensitive data. It assesses attack vectors, including physical attacks, side-channel attacks (e.g., power [61, 85, 86], electromagnetic emission [87], and timing [88]), and software-exploitable hardware attacks [89, 90], and identifies vulnerabilities within the hardware design. This process outlines potential attack scenarios and establishes security requirements for hardware systems. For example, we can explore a particular threat scenario that involves the injection of malicious logic into a design [91]. In this scenario, the threat actor is an employee within a company who possesses database access to the design. This employee exploits their authorized access to introduce harmful code or vulnerabilities into the hardware design or its configuration files. These insider-driven attacks have the potential to establish a hidden entry point that could expose valuable, sensitive information categorized as the assets of the hardware IPs. This threat modeling process helps to outline different attack scenarios or sequences that an attacker might employ to execute his intention. Ultimately, a well-crafted threat model provides the security requisites for any hardware system subject to scrutiny.

*3) Challenges in Security Verification:* The verification process in chip design is a major bottleneck, taking up over 70% of resources and time [92]. Recently, a noticeable increase has also been observed in the instances where the verification process has contributed more than 80% of the total duration of the project [64]. Ensuring secure and compliant SoCs for diverse applications becomes even more challenging, necessitating the identification and addressing of security vulnerabilities in pre-silicon stages. These challenges include the globalized nature of the development lifecycle, complex interactions among hardware, software, and firmware layers leading to unforeseen threats, increased design complexity that limits verification coverage, and the lack of standardized benchmarks for comparing emerging verification techniques. Currently, security verification and validation suffer from limited success due to inadequate prioritization of security in design, lack of suitable threat models and vulnerability databases, and reliance on mostly manual and ad-hoc security analysis and mitigation methods. The state-of-the-art verification methods for security focus on assertion-based security property verification [93], information flow tracking [94], fuzz testing [95], concolic testing [96], penetration testing [97], and AI-based hardware verification [98]. The following discussion describes recent trends in hardware design verification approaches.

1) **Assertion-based Security Property Verification:** Assertion-based security property verification is a vital method for expediting the validation process of hardware designs. It involves incorporating security-related logical statements, or property-based assertions, into the design of a system to formally define security requirements and constraints during development and testing [93, 99, 100]. This technique addresses a key challenge in hardware validation: the limited observability of designs during testing. Observability deals with the capacity to monitor different states within the design. By integrating these properties, we can enhance the observability of design components, enabling the detection of anomalous behavior during simulation. Moreover, assertions based on these properties can identify errors during simulation, significantly reducing the time and resources required for verification. Two widely used assertion specification languages for this purpose are System Verilog Assertions (SVA) [101] and Property Specification Language (PSL) [102], typically applied at the RTL abstraction level of the design. Automated assertion generation techniques have gained popularity to streamline the process of generating assertions and reduce manual effort [103–105]. However, a key challenge with these methods is ensuring the functional accuracy and coverage of the generated assertions. Additionally, these generated assertions operate at the bit level, which can noticeably extend simulation runtimes.

2) **Information Flow Tracking:** Information flow tracking (IFT) within hardware verification is a method utilized for monitoring and regulating the transfer of data or signals within a hardware design or system. Its primary application lies in the domain of security and safety verification, ensuring that sensitive or critical information is managed appropriately and does not unintentionally leak or lead to undesired behavior. IFT effectively identifies and mitigates hardware vulnerabilities, including issues such as timing side channels and hardware Trojans [106, 107]. The most commonly employed IFT techniques include SecVerilog [106], Sapper [108], Caisson [109], and VeriCoqIFT [107]. While these approaches excel at detecting hardware bugs associated with information leakage, they are limited by certain problems, such as the need to learn a new language, the requirement for manual annotation, and the inability to distinguish between implicit and explicit information flows. The RTLLIFT approach, as detailed in [110], addresses these challenges by operating directly within existing HDLs and by enabling the differentiation between implicit and explicit flows. Nonetheless, this approach demonstrates limitations in terms of performance when dealing with complex designs, compelling designers to make trade-offs between precision and computational complexity. Additionally, these approaches are not comprehensive enough to detect most of the security vulnerabilities.

3) **Fuzzing:** It is a popular testing method in software and has recently attracted a lot of attention in the hardware security verification domain. "Fuzz testing" or "fuzzing" denotes a method involving randomized testing of software programs to identify irregularities and weaknesses [111]. Fuzzing typically involves an automated or semi-automated process designed to check a wide range of predefined (instrumented) scenarios involving invalid inputs, with the aim of triggering any existing vulnerabilities within a program. In the hardware domain, fuzzing has been introduced primarily as a solution to address the scalability challenges associated with formal verification methods [112]. Fuzzing can be classified into black-box fuzzing, gray-box fuzzing, and white-box fuzzing based on the information available during the verification phase. Black-box fuzzing relies on the design specifications and is effective for designs with limited information about internal signals [113]. White-box fuzzing is utilized when the design information is completely available [114], and gray-box fuzzing is a hybrid framework to use the best of both white-box and black-box fuzzing techniques [115]. However, these solutions suffer from several drawbacks, such as limited vulnerability coverage, low accuracy due to limited visibility, and poorly defined coverage metrics [95].

4) **Penetration Testing:** Penetration testing involves actively simulating potential attacks to assess the security of a hardware system or device, aiming to discover vulnerabilities and weaknesses [116]. Its primary goal is to evaluate how well hardware components and systems can withstand potential threats and unauthorized accesses. Similar to fuzzing, penetration testing can adopt black-box, white-box, or gray-box approaches based on the specific threat and available resources [117]. This method comprises various stages, ranging from assessing hardware design to exploiting a specific vulnerability that needs resolution. In contrast to random test patterns, penetration testing relies on precise information about security properties, vulnerabilities, and established threat models. However, applying penetration testing in the hardware domain poses greater challenges compared to the software domain because hardware vulnerabilities are more diverse, requiring distinct strategies for different vulnerabilities in each targeted penetration testing scenario [67].

5) **Concolic Testing:** Concolic testing, as an automated test vector generation approach, combines concrete execution with symbolic execution (concolic) to analyze and validate the behavior of a system [118, 119]. This method blends the execution of a program or hardware design, involving actual (concrete) input values, with abstract symbolic representations of input values. The primary objective of Concolic testing is to systematically explore different execution paths within a program or hardware design. By incorporating symbolic inputs, it can simultaneously investigate multiple paths, including those that might be challenging to reach using traditional testing methods. Concolic testing can be resource-intensive, particularly when applied to sizable and complex hardware designs or software programs. In recent times, Concolic testing has found application in hardware security verification, serving purposes such as the detection of hardware Trojans [119], the identification of bugs within the CPU core of SoC [120], and firmware validation [121]. Unfortunately, these methods either only work for a certain part of the SoC due to scalability issues or are limited to detecting a few hardware vulnerabilities.

6) **AI-based Verification:** Machine learning (ML) and deep learning (DL) are emerging tools that have recently gained significant attention within the field of hardware verification. ML techniques find applications in diverse verification processes, including the generation of challenging test cases that are hard to achieve and the validation of test results to enhance coverage. Current research trends in this area include creating constraint-random test vectors through supervised and reinforcement learning [122], fine-tuning decision-making procedures for SAT solvers [98], the identification of hardware Trojans [123], and in-depth debugging of system failure analysis [124].While the prospects are undoubtedly promising, the integration of AI into hardware verification is not without its set of challenges. These hurdles stem from the following factors

- Design dependency: A critical issue often overlooked is the design dependency of AI-based solutions. Due to the lack of a vast and rigorous dataset that involves all corner cases and types of designs, these solutions are often tailored to specific designs and lack adaptability. This limitation hinders their applicability across diverse hardware designs, making them less versatile and effective in a broader context

- Data management: There exists a scarcity of datasets in the hardware security domain that encompass all potential scenarios and corner cases. This data scarcity can lead to the underperformance of ML models that cannot fully optimize the verification process. Lack of benchmark is also a critical challenge for proper evaluation of AI-driven security verification methods.

- Scalability and Efficiency: AI-based solutions are resource-intensive, raising concerns about scalability and efficiency. The time required to train and validate models can be significant, impacting the verification timeline.

- Feature selection and objective function design: There exists a difficulty in identifying and selecting the most relevant features and designing appropriate objective functions for security verification tasks.

In summary, existing verification methods require significant manual labor for security verification, and also suffer from the problems of adaptability and scalability. A notable issue lies in the scarcity of reliable databases for the development of effective techniques and proper evaluation of performance. LLMs hold the potential to introduce creative solutions to address these prevailing challenges in hardware security verification. With proper prompt engineering, LLMs can prove to be highly useful in identifying and mitigating vulnerabilities for complex hardware designs that can reduce a lot of manual effort. Furthermore, the inferential capabilities of LLMs can be harnessed to construct comprehensive databases in the domain of hardware security, which can help solve the problem of lack of data and adaptability in the existing approaches. The incorporation of LLMs can enhance the precision, efficiency, and adaptability of hardware security verification, marking a significant stride toward overcoming existing limitations.

## III. LLM IN SoC SECURITY VALIDATION

In the preceding sections, we thoroughly explored contemporary SoC security validation solutions and their limitations. Building on that foundation, this section ventures into the captivating intersection of LLMs and SoC security. The synergy between LLM and SoC is a pivotal junction where computational linguistics seamlessly integrates with hardware security. We begin with a thorough survey of the established LLMs to offer insights into their learning settings, evolution, architectures, capabilities, and current state. After setting this foundation, our narrative progresses to a deeper exploration of GPTs, the specialized models designed for coding tasks, and LLM chatbots. Later, we explore the art of prompt engineering, highlighting its significance in refining user-model interactions for security validation. We then transition to discussions of the APIs associated with LLMs. Afterward, we describe the importance of fidelity checking in LLM-based solutions. Concluding the section, we turn our attention to the existing works adopting LLM for software and hardware security solutions.

### A. Preliminaries

At first, we narrate three basic concepts, namely learning paradigms, model architectures, and control parameters, that are necessary to understand how LLMs can be incorporated into SoC security validation.

*1) Learning Paradigms in LLM:* As these models evolve, various learning paradigms have been developed to optimize their performance, each with its own set of advantages and challenges. Table I summarizes these methods in terms of data requirement, cost, complexity, and also discusses their role in the context of SoC security. Here, we discuss these learning paradigms and also make our observations in the context of SoC security.

*a) Pre-training:* Pre-training serves as the foundation of the LLM learning pipeline. In this phase, the model is trained on a vast corpus of text data to predict the next word in a sentence. This process allows the model to grasp the intricacies of language, from the basic constructs of grammar to the more abstract concepts of context and semantics. However, while pre-training equips the model with a broad understanding of language, it does not necessarily make the model an expert in specific domains. These pre-trained models can act as foundation models that can be further fine-tuned and adapted to specific domains. BERT [126], GPT-3 [17], LLama-1 [127] are some of these widely used pre-trained models to be named.

**In Context of SoC Security:** Unfortunately, very few existing pre-trained LLMs used HDL codes in their training corpus which narrows the scope of performing security tasks with LLMs. Pre-training an LLM on a large amount of HDL codes, highly specialized with a focus on security-related tasks, can provide a fundamental understanding of hardware security principles. However, the feasibility of this process in an academic setting is challenging primarily because of its significant computational resources, leading to high costs. There is also a scarcity of a large amount of rich HDL codes to be used for the training. Furthermore, the pre-training process has inherent limitations. The nature of the pre-training data used, typically historical text data, means the model may not be equipped to handle emerging, real-time security threats. This could lead to the model having a potential knowledge gap, as it may not be able to respond to the most recent security threats that it has not been trained on.

*b) Fine-Tuning (FT):* Pre-trained models in general have a broad understanding of language and context. However, they often require specialized knowledge to excel in specific domains. This is where fine-tuning steps in, refining the parameters of the model to enhance its performance on specialized tasks. Fine-tuning to the LLM can be executed in various ways: supervised fine-tuning, instruction tuning, and reinforcement learning with human feedback (RLHF) [128]. LHF for LLMs combines reinforcement learning and human feedback to fine-tune models. Initially trained on human responses, the model is later optimized using a reward model derived from user feedback. This approach enhances accuracy, reduces biases, and has notably improved models like InstructGPT [129] and ChatGPT [130]. The cost of fine-tuning is typically less than pre-training but still requires significant resources for optimal results This has led to a growing interest in parameter-efficient fine-tuning methods [131].

**In Context of SoC Security:** In relation to SoC security, fine-tuning can have a powerful impact. It can adeptly handle a range of security tasks, from vulnerability insertion and identification to mitigation and security policy generation. However, realizing this potential is not without its challenges. The foremost among these is the significant cost associated with the fine-tuning process. Furthermore, post fine-tuning, models encounter the "knowledge cut-off" dilemma, rendering them oblivious to threats or vulnerabilities that emerge after their last update. This often leads to the need for recurrent fine-tuning sessions. For example, as of today, the number of hardware CWEs listed by The MITRE Corporation [80] is 104. This number is not static—it increases as new vulnerabilities are discovered. Consider a model that has been fine-tuned to recognize and mitigate the existing 104 CWEs. As the list expands with the discovery of new vulnerabilities, the "knowledge cut-off" of the model renders it ineffective against

| Learning Setting | Data Requirement [125] | Cost | Complexity of task | Prospect of SoC Security |
|---|---|---|---|---|
| Pre-training | >1 trillion tokens | Very High | High | Not feasible due to high cost and lack of data |
| Fine-Tuning (Supervised Fine-Tuning) | 10-100k prompt-response | High | Medium | Good for all tasks Limited training data & benchmark. |
| Fine-Tuning (RLHF) | Step 1: 100k-1M examples, Step 2: 10-100k prompts | High | Very High | Good for vulnerability insertion, detection & mitigation. |
| In-Context Learning (ICL) | 2-5 examples | Low | Low | Decent at all tasks. Lack of long term context |
| Retrieval Augmented In-Context Learning (RA-ICL) | 2-5 examples + knowledge source | Medium | High | Decent at vulnerability detection & mitigation. |

these newly identified threats. It is confined to the knowledge it was last updated with, requiring regular fine-tuning to maintain its effectiveness and relevance. Currently, in the domain of SoC security, there exists a potential lack of large datasets for different security tasks which can be for effective fine-tuning. The absence of standardized benchmarks makes objective evaluation challenging for a fine-tuned model.

*Prospect 1:* Fine-tuning can be strategically employed to adapt LLM to specific security tasks, seamlessly bridging the gap between generalized knowledge and domain-specific expertise.

*Challenge 1:* In training LLM for SoC security, a key challenge arises from the limited availability of domain-specific training data and benchmarks, both crucial for refining and validating model performance in specialized security scenarios.

*c) In-Context Learning (ICL):* In-context learning (ICL), as highlighted in GPT-3 [17], is one of the game-changing capabilities of LLM. ICL, including zero-shot, one-shot, and few-shot learning, allows GPT models to adapt to new tasks bypassing the need for traditional fine-tuning by generating responses based on the instructions supplemented with or without examples. The benefits of ICL are numerous. Firstly, it is adept at broadening its functionality to new tasks using a limited number of examples. Secondly, ICL reduces the requirement for extensive computational resources, making it a more efficient learning approach. Lastly, ICL navigates issues such as overfitting [132] and frequency shock [133], commonly encountered in traditional learning methodologies. In-context learning (ICL) offers benefits but can falter with certain queries due to prompt complexity, quality, and demonstration distribution. These factors play a crucial role in shaping the learning and generalization capabilities of the model.

*In Context of SoC Security:* In the In-context learning, as applied to the SoC security domain, offers a unique blend of adaptability and specificity. By leveraging the ability of the model to understand and respond based on the provided context, it can offer solutions tailored to specific security challenges. This is particularly beneficial for dynamic security analysis, where the context can vary based on the design, threat landscape, or specific security protocols in place. However, one of the primary challenges is the ability of the model to maintain a long-term context. In SoC security, scenarios often span complicated designs and complex threat landscapes, requiring a deep and prolonged understanding of the context. If the model struggles to retain or comprehend this extended context, it might offer solutions that are fragmented or lack depth. This limitation could inhibit the model from effectively handling ongoing security scenarios, especially those that require a holistic understanding of the system, its vulnerabilities, and potential mitigation strategies.

*Achievement 1:* ICL in LLMs has revolutionized adaptability, enabling the model to tackle new tasks without traditional fine-tuning, proving particularly beneficial in dynamic security analysis for SoC security.

*Challenge 2:* ICL can struggle with maintaining a long-term context, especially in complex scenarios like SoC security, leading to potentially fragmented solutions that may not effectively address intricate, ongoing security issues.

*d) Retrieval-Augmented In-context Learning (RA-ICL):* Pre-trained GPT model does not have access to external knowledge. In order to incorporate up-to-date information that the LLM has not seen during training, it must be retrained. As previously described, ICL also relies solely on the information available during its last training update and may lack the most current information. Retraining these models to update their knowledge is an option but comes with significant costs, both in terms of time and resources. In order to tackle these issues, RA-ICL [134, 135] comes in handy. It addresses these issues by grounding the model during generation by conditioning on relevant documents retrieved from an external knowledge source. Retrieval-Augmented Language Modeling (RALM) systems have two components: knowledge retriever and knowledge generator. The knowledge retriever is responsible for sourcing and retrieving relevant information from an extensive external database, ensuring that the most current and pertinent data is accessed. On the other hand, the knowledge generator takes this retrieved information and integrates it into the response generation process, ensuring that the outputs are not only contextually appropriate but also enriched with the latest information. This approach enriches response generation but might also increase computational complexity and costs.

*In Context of SoC Security:* In SoC security, retrieval-augmented in-context learning can allow GPT models to pull relevant security information from external databases, enhancing the comprehensiveness and relevance of the security analysis. In a scenario described in Section III-A1b, RA-ICL can solve the problem of "knowledge cut-offs" in the task of hardware vulnerability detection and mitigation. The knowledge retriever component actively can scan and retrieve new CWE data, ensuring that the responses of the model are informed and current.

***Prospect 2:*** The adoption of RA-ICL promises enhanced real-time adaptability in SoC security solutions, as it is capable of continuously updating its knowledge base with the most current information, ensuring that responses and solutions are always informed, relevant, and up-to-date.

*2) Model Architecture:* Based on the model architecture, we discuss four categories of existing LLMs: decoder-only, encoder-only, encoder-decoder, and sparse models. A detailed discussion of their structure, working principles, training objectives, functions, and role in the context of hardware design and SoC security is given below.

*a) Encoder-Decoder Model:* The most well-known implementation of the encoder-decoder architecture is the transformer, introduced in [136]. This model is a two-part architecture. The encoder processes the input sequence and compresses it into a context or an intermediate representation. The encoder component transforms input tokens into vectors using embeddings and positional encodings, then applies multihead self-attention and feedforward networks. The decoder, starting similarly, incorporates masked self-attention and cross-attention with the output of the encoder, ensuring alignment and preventing future word prediction. Sequence-to-sequence models often use this architecture, where the input and output sequences can be of different lengths. These models shine in tasks where there is a direct and complex transformation between inputs and outputs. Examples like machine translation and text summarization are prototypical, as they require the model to understand the input deeply and generate a coherent and contextually accurate output. BART [5], T5 [137], and UL2 [138] are a few well-known encoder-decoder models to be named.

***In Context of SoC Security:*** The encoder-decoder architecture, renowned for its ability in natural language understanding tasks, exhibits versatility in SoC security. Its two-stage process of encoding the input data and then decoding it to produce an output makes it suitable for tasks that require both comprehension and generation. This model is particularly adept at vulnerability mitigation, where understanding the context (encoder) and generating a solution (decoder) are both crucial. However, while it is also a good fit for tasks like vulnerability insertion, security verification, and assessment, it might not always be the optimal choice when the task leans heavily toward either comprehension or generation

*b) Decoder-Only:* Decoder-only LLMs have established impressive benchmarks in numerous NLP tasks, especially in the generation of free-form text. In a decoder-only model, a sequence is fed into the model, which then directly predicts the next token or word in the sequence. It operates autoregressively, using its generated tokens as context for subsequent predictions. It has two variants: causal decoder and prefix decoder. In a standard decoder (causal decoder), the unidirectional attention masking ensures that a token attends only to previous tokens and itself. Prefix decoders permit bidirectional attention over prefix tokens while maintaining unidirectional attention to generated tokens. These models excel in tasks like dialog and story generation that require deep input understanding and coherent output generation. Decoder-only architectures, such as the GPT series [17, 18], have gained popularity due to their parameter efficiency, simplicity, generalization, and versatility.

***In Context of SoC Security:*** Decoder-only models, known for their strength in unconditional generation tasks, shine in areas of SoC security that are predominantly generative. They are tailored for tasks like vulnerability insertion, security policy and property generation, and testbench generation, where the model needs to produce new content based on a given prompt or context. For tasks demanding a deeper understanding before generation, like vulnerability mitigation, they can still offer decent performance but might not be the primary choice.

*c) Encoder-Only Model:* Encoder-only models process input sequences and output a fixed-size context for each token or the entire sequence. These models are adept at distilling information from input sequences into fixed representations, making them suitable for tasks like classification where the aim is to derive a condensed understanding from the input. These models are referred to as "encoder-only" because they prioritize encoding input sequences into meaningful embeddings. The "decoding" they do is not about generating novel sequences (as with autoregressive models), but rather about producing specific outputs from the learned embeddings, such as masked token predictions during pretraining. BERT (Bidirectional Encoder Representations of Transformers) [126] developed by Google and its variants: RoBERTa [139], DistilBERT [140], etc are popular examples. These models have lost popularity in recent times.

***In Context of SoC Security:*** Encoder-only models, with their inherent design to understand and represent data, align well with tasks that require profound analysis. In the SoC security landscape, they are best suited for tasks like security verification and assessment, which demand an in-depth comprehension of the given data without extensive generation. However, when the task requires subsequent generative actions based on the understood context, encoder-only models might not be the ideal choice.

*d) Sparse Model:* Sparse models, particularly those rooted in the mixture-of-experts (MoE) [141] paradigm, are on the leading edge of LLM architecture innovation. The core idea behind them is to selectively activate only a subset of model parameters for each input, ensuring a more efficient computation without compromising the capacity of the model. In a typical dense neural network, every input engages the full spectrum of the parameters of the model. In contrast, sparse models, and especially those based on the MoE framework, allocate only a specific set of 'expert' parameters tailored for each input. The Switch Transformer [142] and GShard [143] are some of the notable models in this domain. They underline how sparse activation mechanisms can achieve, and sometimes surpass, the efficacy of dense counterparts but with greater computational efficiency.

TABLE II
Performance of different model architectures across various security tasks. The "Best" column indicates the model type that is most suited for the task, offering optimal performance. The "Decent" column lists model types that can perform the task reasonably well but might not be the optimal choice. The "Poor" column indicates model types that are least suited for the task and might not deliver satisfactory results.

| Security Task | Performance | | |
|---|---|---|---|
| | Best | Decent | Poor |
| Vulnerability Insertion | Decoder-only | Encoder-decoder | Encoder-only |
| Security Verification | Encoder-only | Encoder-decoder Decoder-only | - |
| Security Assessment | Encoder-only | Encoder-decoder Decoder-only | - |
| Security Policy & Property Generation | Decoder-only | Encoder-decoder | Encoder-only |
| Testbench Generation | Decoder-only | Encoder-decoder | Encoder-only |
| Vulnerability Mitigation | Encoder-decoder | Decoder-only | Encoder-only |

*In Context of SoC Security:* In the context of SoC security, where data can be both vast and intricate, sparse models can provide an optimal balance between computational efficiency and task-specific precision

The potential roles of different model architectures in SoC security tasks are summarized in Table II.

*Prospect 3:* By selecting the optimal model architecture, one can achieve greater precision in executing specific security tasks.

*3) Control Parameters:* One of the benefits of LLM or GPT over other traditional deep learning approaches is that the nature of the generated output can be controlled through several parameters [144]: temperature, top_p, presence_penalty and frequency_penalty. They do not influence the learning process of the model during training. Instead, they are parameters used at inference time when generating outputs from the trained model. A brief definition of these parameters is given below.

- Temperature: It is a parameter that controls the randomness of predictions in the output sequence. When applied to the output probabilities, a high temperature makes the probability distribution more uniform leading to more diverse and potentially more creative output tokens, while a low temperature makes the distribution the distribution more focused resulting in more deterministic and potentially less creative output tokens.
- top_p: It is an alternative to the temperature used in nucleus sampling, where with this method tokens are selected from the smallest set of top-ranked tokens whose cumulative probability exceeds a certain threshold (p). Lower top_p values result in a smaller, more focused selection of tokens, which leads to less diverse output.

Indeed, temperature and top_p serve as crucial parameters in shaping the generation of text or code by LLMs, each influencing the randomness and diversity of the output. For instance, when the goal is to generate alternative versions of a given code segment, a high temperature or top_p can lead to a range of diverse and innovative solutions. On the contrary, in generating code following a specific pattern or protocol, a lower temperature or top_p value can help ensure that the resulting code adheres to the desired structure.

*In Context of SoC Security:* These parameters offer promising avenues for exploration and optimization in the context of SoC security. Their manipulation can impact various tasks related to SoC security, including design generation, vulnerability insertion, security rule and property creation, assertion generation, security assessment, and vulnerability mitigation. However, the optimal setting of these parameters may vary significantly depending on the specific task and the desired balance between creativity and adherence to established patterns or rules. There is a need for a comprehensive investigation of how these parameters should be selected for a particular security task in order to fully leverage their potential in enhancing SoC security.

*B. Existing LLMs*

The introduction of the transformer [136] model has had a profound impact on NLP, leading to significant improvements in performance across a wide range of tasks. The transformer architecture is the foundation for several state-of-the-art models such as BERT [126], GPTs [17, 18, 145, 146]. It is very important to understand the current situation of the development of LLMs. In this light, Table III provides a crucial snapshot of the rapid evolution of LLMs. It offers a consolidated view of various models, highlighting their architectures, training schemes, and unique features. This comprehensive overview will help in making informed decisions about model selection and deployment. For example, by observing trends in model sizes and training techniques, one can anticipate future directions in AI research. We make the following key observations from the table.

- **Diverse Developers:** While technology behemoths such as OpenAI, Google, and Microsoft lead in terms of the number and variety of LLMs, the emergence of models from organizations like EleutheAI and BigScience showcases the democratization of AI research. The collaboration between Microsoft and NVIDIA for MT-NLG signifies that strategic partnerships can lead to advancements in LLMs.
- **Architectural Variations:** Decoder-only models seem to dominate the LLM space, indicating their success in tasks related to language generation and completion. The diversity in architectures, with Encoder-Only, Encoder-Decoder, and MoE, indicates ongoing experiments in capturing different facets of language understanding and generation.

TABLE III

A DETAILED OVERVIEW OF EXISTING LLMS SHOWCASING THEIR DEVELOPMENT LINEAGE, ARCHITECTURE, AND UNIQUE CHARACTERISTICS. THE TABLE HIGHLIGHTS THE DEVELOPER, ARCHITECTURE TYPE, BASE MODEL, NUMBER OF PARAMETERS, TRAINING SCHEMES, PUBLIC AVAILABILITY, AND SPECIFIC FEATURES.

| Developer | Model | Architecture | Base Model | # parameters (B) | Training Scheme | Publicly Availability | Remarks |
|---|---|---|---|---|---|---|---|
| Open AI | GPT-1 [145] | Decoder-Only | - | 0.117 | Pre-train + SFT | No | Fine-tuning on specific task |
| | GPT-2 [146] | Decoder-Only | - | 1.5 | Pre-train | No | Multi-tasking and zero-shot setting |
| | Codex [35] | Decoder-Only | GPT-3 | 12 | FT | No | Code generation and other coding tasks |
| | GPT-3 [17] | Decoder-Only | - | 175 | Pre-train + ICL | No | Few-shot learning setting |
| | WebGPT [147] | Decoder-Only | GPT-3 | 175 | FT | No | Long-form question-answering |
| | InstructGPT [129] | Decoder-Only | GPT-3 | 175 | RLHF | No | Fine-tuning with human feedback |
| | GPT-4 [18] | Decoder-Only | - | - | IT +RLHF | No | Remarkable improvement in complex task |
| Meta | RoBERTa [139] | Encoder-Only | BERT | 0.125 (base) 0.355 (large) | Pre-training | Yes | Based on BERT but with different hyperparameter choices |
| | BART [5] | Encoder-Decoder | - | 0.140 (base) 0.400 (large) | Pre-training + FT | No | Particularly effective when fine-tuned for text generation |
| | NLLB [148] | MoE | - | 54.5 | Pre-training | Yes | Machine translation over 202 languages |
| | LLama-1 [127] | Decoder-Only | - | 65 | Pre-training | Yes | Open language model |
| | LLama-2 [149] | Decoder-Only | - | 7-70 | Pre-training | Yes | LLama-2-Chat though SFT and RLHF on LLama-2 |
| | Galactica [150] | Decoder-Only | - | 120 | Pre-training | Yes | Trained on corpus of scientific knowledge |
| | OPT [151] | Decoder-Only | - | 175 | Pre-training | Yes | Open pre-trained transformer |
| | OPT-IML [152] | Decoder-Only | OPT | 30, 175 | IT | Yes | Instruction-tuning for generalisation |
| Google Research | BERT [126] | Encoder Only | - | 0.110 (base) 0.340 (large) | Pre-training | Yes | Good for language understanding tasks |
| | XLNet [153] | Encoder-Only | Transformer-XL | 0.110 (base) 0.340 (large) | Pre-training | Yes | Combination of ideas of BERT and traditional autoregressive model |
| | AlBERT [154] | Encoder-Only | BERT | 0.011 | Pre-training | Yes | Improves parameter-efficiency |
| | T5 [137] | Encoder-Decoder | - | 11 | Pre-training + FT | Yes | Transfer learning for NLP |
| | UL2 [138] | Encoder-Decoder | - | 20 | Pre-training | Yes | Different training objective |
| | mT5 [155] | Encoder-Decoder | T5 | 13 | Pre-training + FT | Yes | Multilingual variant of T5 |
| | Flan-T5 [156] | Encoder-Decoder | T5 | 11 | FT | Yes | Instruction fine-tuning on T5 |
| | Gshard [143] | MoE | - | 600 | Pre-training | No | Scaling up multilingual translation |
| | LamDA [157] | Decoder-Only | - | 137 | FT | No | Specialized for dialog |
| | FLAN [158] | Decoder-Only | LamDA-PT | 137 | FT | No | Uses instruction fine-tuning |
| | PaLM [19] | Prefix Decoder | - | 540 | - | No | Pipeline free training on large scale |
| | Minerva [159] | Decoder-Only | PaLM | 540 | Pre-training + FT | No | LLM for quantitative reasoning problem, especially math problem |
| | Flan-PaLM [156] | Prefix Decoder | PaLM | 540 | FT | No | Uses different instruction fine-tuning |
| | PaLM 2 [160] | Prefix Decoder | - | - | Pre-training | No | Multilingual & reasoning capabilities Smaller in size but compute-efficient |
| DeepMind | AlphaCode [34] | Encoder-Decoder | - | 41 | FT | No | System for code generation |
| | Chinchilla [161] | Decoder-Only | - | 70 | Pre-training | No | Compute-optimal model over Gopher |
| | Sparrow [162] | Decoder-Only | Chinchilla | 70 | SFT+RLHF | No | Information-seeking dialogue agent |
| | Gopher [163] | Decoder-Only | - | 280 | Pre-training | No | Compares Dialogue-Prompted Gopher and Dialogue-Tuned Gopher |
| Microsoft | DialogGPT [164] | Decoder-Only | GPT-2 | 0.117,0.345, 0.762 | Pre-training | Yes | Generates conversational responses |
| | DeBERTa [165] | Encoder-Only | - | 0.1 (base) | Pre-training | Yes | Improves BERT and RoBERTa |
| | DeBERTaV3 [166] | Encoder-Only | DeBERTa | 0.086 | Pre-training +FT | Yes | Improves DeBERTa |
| Microsoft & NVIDIA | MT-NLG [167] | Decoder-Only | - | 530 | Pre-training | No | Comparable to GPT-3 but larger |
| EleutheAI | GPT-Neo [168] | Decoder-Only | GPT-3 | 1.3 | Pre-training | Yes | Not fine-tuned for downstream tasks |
| | GPT-J 6B [169] | Decoder-Only | - | 6 | Pre-training | Yes | Not fine-tuned for downstream tasks |
| | Pythia [170] | Decoder-Only | - | 12 | Pre-training | Yes | Not fine-tuned for downstream tasks |
| AI21 Labs | Jurassic-1 [171] | Decoder-Only | - | 178 | Pre-training | Yes | Comparable to GPT-3 |
| | Jurassic-2 [172] | Decoder-Only | - | - | - | No | Capable of composing human-like text |
| Yandex | YaLM [173] | Decoder-Only | Megatron-LM | 100 | Pre-training | Yes | Focuses Russian and English text |
| TII | Falcon [174] | Decoder-Only | - | 40 | Pre-training | Yes | Fundamentally based on GPT-3 Falcon-40B-instruct is fine-tuned o |
| BigScience | BLOOM [175] | Decoder-Only | - | 176 | Pre-training | Yes | Trained on 46 languages and 13 programming languages |
| | BLOOMZ | Decoder-Only | BLOOM | 176 | FT | Yes | Zero-shot task generalization abilities |
| Others | GLM [176] | Decoder-Only | - | 130 | Pre-training | Yes | Bilingual pre-trained LLM |

- **Training Schemes:** Evolving training methods, including RLHF, highlight the industry's push towards refining model outputs using human feedback.
- **Model Size:** As technology advances, there is a clear trend towards building larger models. However, models like PaLM 2 are emphasized as compute-efficient, highlighting the importance of balancing size with practicality and resource constraints.

- **Public Availability:** Some models, especially those developed by tech giants like OpenAI, remain proprietary. Ont he other hand, openly available models, like LLama by Meta or BERT by Google, have spurred a plethora of research, with many in the community building upon or fine-tuning these models for specific tasks.

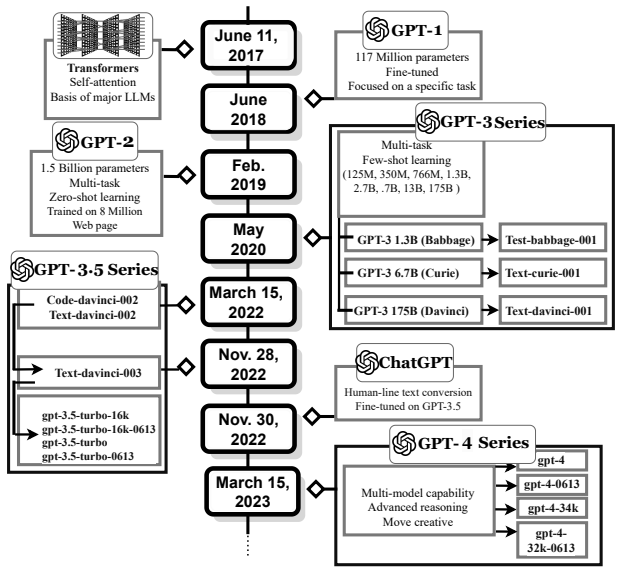

Fig. 3. Timeline for the evolution of GPTs.

## C. Evolution of GPT

The Transformer [136] model revolutionized NLP by introducing attention mechanisms, eliminating the need for recurrent or convolutional layers. This architecture paved the way for the OpenAI GPT series. GPT-1 [145], launched in 2018, was a breakthrough in language modeling with 117 million parameters. GPT-2 [146], with 1.5 billion parameters, improved coherence in text generation, but had concerns about misleading information. Later, GPT-3 [17], introduced in 2020 with a massive 175 billion parameters, achieved near-human text generation and exhibited "few-shot learning," although it still had some limitations. Subsequent models, GPT-3.5 and GPT-4, further refined these capabilities, addressing previous limitations and setting new benchmarks in the field. The GPT-3.5 series includes models such as 'gpt-3.5-turbo' and its variants, with token limits up to 16,384. Some of these models are chat-optimized, while others, like 'text-davinci-003' and 'text-davinci-002', are superior in language tasks. The Codex model, 'code-davinci-002', is optimized for code completion. Lastly, the GPT-3 series also encompasses models like 'davinci', 'curie', 'babbage', and 'ada', all with a token limit of 2,049. These models vary in efficiency, speed, and cost, with some being also available for fine-tuning. Within the GPT-4 series, there are models like 'gpt-4', 'gpt-4-0613', 'gpt-4-32k', and 'gpt-4-32k-0613', all having a maximum token limit ranging from 8,192 to 32,768. These models are advanced, with some versions optimized for chat, and others offering extended context lengths. A timeline of the evolution of the GPTs is shown in Figure 3.

***Achievement 2:*** GPT models, especially GPT-4, stand superior to other counterparts due to their unparalleled understanding of both natural language and codes, combined with their emerging reasoning abilities, marking a remarkable shift in the domain.

***Challenge 3:*** High cost associated with GPT models hinders their widespread adoption for large-scale hardware design.

***Challenge 4:*** The proprietary designs of GPT models are not open-sourced, limiting researchers and developers from understanding their intricacies and functionalities to conducting fundamental research in the domain.

## D. LLM for Coding Task

LLMs have shown impressive capabilities in coding tasks, bridging the gap between natural language processing and software development. Over the past few years, they have emerged as invaluable tools for developers, streamlining various stages of the software life cycle. From code generation, defect detection, and auto-documentation, to assisting in debugging and even predicting potential software vulnerabilities, LLMs have transformed traditional coding paradigms. Various fine-tuned LLMs specially dedicated to coding tasks have been released by fine-tuning pre-trained models. Table IV lists all such code LLMs developed in recent years. Analyzing the table, a few key observations emerge. Firstly, the LLMs have predominantly been developed with a focus on mainstream programming languages like Python. This emphasis on Python and similar languages is understandable given their widespread use in software development and data science. However, this has inadvertently led to a gap in the landscape of LLMs specifically fine-tuned for HDL such as Verilog and VHDL. Second, a significant number of these models have undergone fine-tuning, underscoring the importance of domain-specific training for coding tasks.

***In Context of SoC Security:*** There is a noticeable scarcity of fine-tuned LLMs in Verilog, indicating a potential gap in the handling of HDL. However, given that these models already possess a foundational understanding of coding constructs and logic, there is a promising avenue to further fine-tune them for HDLs, leveraging their inherent coding expertise. this potential fine-tuning becomes even more significant. HDLs play a pivotal role in designing and verifying hardware components, making them crucial in ensuring the security and reliability of integrated circuits and systems. By adapting code-centric LLMs to better understand and generate HDL code, we can harness their capabilities to detect vulnerabilities, suggest mitigations, and even aid in the design of secure hardware components, thus bolstering the overall security posture of SoC designs.

***Achievement 3:*** LLMs have made remarkable strides in understanding and generating mainstream software programming languages, transforming traditional software development paradigms.

***Challenge 5:*** The current landscape of LLMs is predominantly tailored for mainstream programming languages, leading to a gap in specialized models for HDLs.

***Prospect 4:*** There is a promising avenue to fine-tune existing code-centric LLMs, which are proficient in mainstream programming languages, for HDLs to bolster SoC security.

## E. LLM for Chatbot

The evolution and diversity in the development of LLMs for chatbot applications are clearly evident from Table V.

TABLE IV
A COMPREHENSIVE LIST OF LLMS SPECIFICALLY DEVELOPED FOR CODING TASKS. THE TABLE PROVIDES INSIGHTS INTO THE EVOLUTION AND DIVERSITY OF CODE-CENTRIC LLMS, EMPHASIZING THE VARIED APPROACHES AND INNOVATIONS TAILORED FOR CODE GENERATION, UNDERSTANDING, AND OTHER CODE-RELATED TASKS.

| Code LLM | Model Architecture | Parameters | Training | Base Model | FT for HDL? | NL+ Code | Open/ Closed | Task Paradigms |
|---|---|---|---|---|---|---|---|---|
| AlphaCode [34] | Encoder-Decoder | 1B (base) | Pre-training + FT | - | No | Code | Closed | Code generation |
| CodeBERT [42] | Encoder-Only | 125M | Pre-trained | RoBERTa-base | No | NL+Code | Open | Code search and code-to-documentation generation |
| CodeGeex [41] | Decoder-Only | 13B | Pre-training + FT + Few-shot | - | No | NL+Code | Open | Pre-trained, fine-tuned & few-shot prompted model performs code generation, translation & explanation, respectively |
| CodeGen-Mono [36] | Decoder-Only | 350M-16.1B | Pre-Training + FT | CodeGen-Multi | No | NL+Code | Open | Code generation |
| CodeGen-Multi [36] | Decoder-Only | 350M-16.1B | Pre-Training + FT | CodeGen-NL | No | NL+Code | Open | Code generation |
| CodeGen2.5-Multi [37] | - | 7B | Pre-Training + FT | - | No | NL+Code | Open | Code generation and in-filling |
| Code LLama [177] | Decoder-Only | 7, 13 & 34 | FT | LLama-2 | No | NL+Code | Open | Code generation and understanding |
| Code LLama-Python [177] | Decoder-Only | 7, 13 & 34 | FT | Code LLama | No | Python Code | Open | Specialized for Python |
| Code LLama-Instruct [177] | Decoder-Only | 7, 13 & 34 | IT | Code LLama | No | Code | Open | Code generation and understanding |
| CodeT5 [178] | Encoder-Decoder | 60M, 220M | Pre-Training + FT | T5 | No | Code | Open | Code understanding and generation tasks |
| CodeT5+ [179] | Encoder-Decoder | 220M-16B | Pre-Training | T5/ CodeGen-Mono | No | Code | Open | Code understanding (retrieval, defect and clone detection) and summarization, generation |
| Codex [35] | Decoder-Only | 12B | FT | GPT-3 | No | Code | Closed | Generation and others |
| Incoder [180] | MoE | 6.7 B | Pre-training | FairSeq | No | NL+ code | Open | Generation, masking and infilling |
| JuPyT5 [181] | Encoder-Decoder | 350 M | Pre-Training | BART | No | Code | Closed | Code in-filling |
| LongCoder [182] | Long-Range Transformer | 150M | Pre-training | UnixCoder | No | Code | Closed | Code complition |
| PanGu-Coder [183] | Decoder-Only | 317M 2.6B | Pre-training | PANGU- | No | Code | Closed | Text-to-code generation |
| PanGu-Coder2 [43] | Decoder-Only | 15B | RRTF (Evol-Instruct) | - | No | Code | Closed | Code generation |
| PLBART [184] | Encoder-Decoder | 140M | Pre-training +FT | BART | No | NL+ Code | Closed | Summarization, generation, translation and classification |
| PolyCoder [185] | Decoder-Only | 2.7B | Pre-training | GPT-2 | No | Code | Open | Code generation |
| SantaCoder [186] | Decoder-Only with FIM & MQA | 1.1 B | Pre-training | - | No | Code | Open | Infilling capabilities |
| StarCoder [39] | Decoder-Only with FIM & MQA | 15.5 B | Pre-training + FT | StarCoder-Base | Yes | Code | Open | Infilling capabilities |
| VeriGen [187] | Decoder-Only | 16B | FT | CodeGen | Yes | Code | Open | Verilog code generation |
| WizardCoder [40] | Decoder-Only with FIM & MQA | 15.5B | FT (Evol-Instruct) | StarCoder | No | Code | Open | Code generation |

OpenAI's development from ChatGPT with GPT-3.5 to GPT-4, released between November 2022 and March 2023, emphasizes the organization's dedication to improving human-like text generation, with the latter version focusing on reasoning and multi-modal capabilities. Meanwhile, Google's 'Bard' and 'Baize', developed around similar timelines, demonstrate their shift from LaMDA to PaLM 2, where Bard stands out for its multi-language support and image-inclusive answers. Notably, Anthropic's Claude 2, with its substantial context length of 100k, highlights a potential trend towards understanding more extensive user inputs. Furthermore, Vicuna's collaborative development by multiple universities showcases the academic interest in LLMs, emphasizing the utility of ChatGPT conversations for fine-tuning. Finally, MosaicML's MPT-Chat presents a significant leap with a 30B parameter size and an impressive 8k context length, which underscores the ever-growing scale and capability of LLMs in the chatbot domain.

***In Context of SoC Security:*** The landscape of chatbots capable of writing HDL codes is still emerging. Among the contenders, ChatGPT stands out as a superior choice, primarily due to its ability to handle extended context lengths. However, even with its advanced capabilities, ChatGPT occasionally falls short when dealing with larger design frameworks, which subsequently impacts its efficiency in other security tasks. Another pressing concern in the SoC security domain is the knowledge cut-off inherent to these models. Given that SoC security is a rapidly evolving field, the static nature of a model's knowledge base can pose challenges. As the domain undergoes continuous advancements, the inability of models like ChatGPT to stay updated in real-time becomes a significant limitation, potentially hindering their applicability in addressing the latest security concerns.

***Challenge 6:*** Due to limited context length, LLM chatbots still fall short for larger design tasks, affecting their efficiency in various security tasks.

***Challenge 7:*** As the domain of SoC security is continuously changing, LLMs might not be up-to-date with the latest developments, potentially impacting their effectiveness in addressing recent security concerns

TABLE V
Overview of prominent LLM Chatbots, highlighting their foundational models, parameter sizes, context lengths, training methodologies, knowledge cut-offs, access, and unique features.

| Chatbot | Base Model | Parameter | Context Length | Training | Knowledge Cut-Off | Access | Remarks |
|---------|-----------|-----------|----------------|----------|-------------------|--------|---------|
| ChtaGPT [130] | GPT-3.5 | - | - | SFT + RLHF | Sep 2021 | Open | Generates human-like text across a wide range of topics |
| | GPT-4 | - | - | SFT+ RLHF | Sep 2021 | Limited | Adds reasoning & multi-modal capability |
| Bard | LaMDA (Initial) PaLM 2 (current) | - | - | FT + RLHF | - | Open | Supports 40+ language Includes image in answer |
| Baize [188] | LaMDA | 7B | 4096 | SFT + SDF | | Open | Generation token limit: 512 |
| Claude 2 | Claude 2 | - | 100k | RLHF | Early 2023 | Limited | Ability to interpret longer input |
| Vicuna [189] | LLaMa | 7, 13, 33B | 2048 | FT with LoRA + SDF with ChatGPT | - | Open | Fine-tuned with 70k shared ChatGPT conversations |
| MPT-Chat [190] | MPT-30B | 30B | 8192 | FT | - | Open | Trained with 8k context length |

## F. API of LLMs

The increasing prominence and capabilities of LLMs have introduced significant computational and memory demands. Furthermore, a significant number of LLMs, especially those developed by OpenAI, are not publicly accessible. Due to such hardware limitations, hardware costs, limited accessibility, and scalability issues, it becomes infeasible to run LLMs locally and requires an alternative method for their use, especially for common use. Here, application programming interfaces (APIs) come in handy. APIs serve as a bridge, enabling users to seamlessly take advantage of the immense capabilities of LLMs without confronting the technical complexities associated with local deployments. APIs offer a more accessible, cost-effective, and efficient means to leverage the capabilities of LLMs. Several companies such as OpenAI, Cohere, AI21 Labs, and Anthropic have already recognized the demand and potential of APIs [191–194], and have thus provided a diverse range of interfaces catering to different applications and computational needs. Knowledge of these APIs is essential to access, understand, and utilize the latest developments in LLMs, enabling seamless integration and exploitation of enhanced capabilities in various applications.

*In Context of SoC Security:* OpenAI APIs are still the most prominent ones to be familiar with in order to exercise SoC security validation tasks. To stay abreast of the latest developments and offerings of APIs by OpenAI, readers are encouraged to consult [195]. A significant hurdle of using such APIs is the comparatively high cost of usage. For instance, the current expense of fine-tuning GPT-3.5-turbo involves a substantial investment; training a model with a 10M token file over 3 epochs costs approximately 240 USD. Moreover, the newest, high-performance models, such as GPT-4, are not yet accessible for fine-tuning, limiting the options for users seeking the most cutting-edge solutions.

## G. Prompt Engineering

Prompt engineering [196] refers to the process of crafting natural language prompts that guide an LLM during inference. It has been observed that prompting techniques can significantly influence the behavior of LLM. This has led to the emergence of several prompting techniques [197–210]. In the subsequent discussion, our focus will be on these prompting methodologies that have been suggested in the previous works, especially for coding and reasoning tasks. We categorize the prompting methods into the following categories.

- **Task Decomposition:** Task decomposition refers to the breaking down of a larger or more complex problem into smaller, more manageable sub-tasks or components. This allows for the easier tackling of each smaller task, which when combined, solves the original problem. The least to the most prompting [197] and the subsequent prompting [198] are two of such methods. The least-to-most prompting [197] first breaks down a complex problem into a series of simpler sub-problems, and after the problem is fully decomposed into sub-problems, these sub-problems are then solved in sequence. Unlike Least-to-most prompting, in the successive prompting [198] method, the decomposition of the question and the answering stages are interleaved. This means that as a sub-problem is identified, it might be immediately solved before moving on to decomposing the next part of the problem.

- **Sequential Reasoning:** Sequential reasoning methods [199–202], with a little difference from task decomposition, follow a chain of logical steps to solve a problem. Chain of Thought (CoT) prompting is a popular multi-step reasoning technique that enhances the complex reasoning abilities of LLMs by guiding them through a series of intermediate reasoning steps, often demonstrated via exemplars. Zero-CoT is a variant of CoT that adds phrases like "Let's think step by step", and the LLM is cued to reason sequentially, even without prior examples. A major limitation of CoT is its efficacy is limited in tasks needing exploration. Tree of Thoughts (ToT) is an advanced version of CoT. It generalizes the CoT technique and enables the exploration of different "thoughts" that act as intermediate steps in problem-solving. Other researchers improved the performance of CoT through self-consistency [211] and greater complexity of reasoning[203].

- **Self-Evaluation and Refinement:** The methods facilitate the LLM to critique, rectify, or refine its own outputs. In such methods, the LLM acts as both an executor and an evaluator. Self-Debugging [206] approach prompts the LLM to debug its predicted code. After code execution, the model generates feedback based on code performance, helping toy and correct errors. In Self-Refine [205], the method generates an initial LLM output which is then refined iteratively based on feedback from the same LLM. Recursive Criticism and Improvement (RCI) [207] is another technique that starts with a zero-shot prompted LLM output. Then, the LLM is asked to identify problems with that output, and based on the detected issues, the LLM generates an updated solution.

- **Intermediate Output Visualization:** This methodS emphasize displaying intermediate reasonings, especially helpful for tasks that require multi-step computations. In [208], the authors prompt transformers to execute multi-step computations by asking them to show intermediate computational results on a "scratchpad".
- **Multi-LLM Collaboration:** Focused on the cooperative use of multiple LLMs, these methods leverage the strengths of different models for better problem solving. For example, [209] uses two fine-tuned LLMs in tandem, one dedicated to selection and the other to inference, fostering a cohesive reasoning process.

*In Context of SoC Security:* SoC security-related tasks, i.e., security vulnerability injection, detection, and countermeasure development are inherently complicated. The traditional prompting approach - typically by framing queries in a standard question-answer format— does not always give the intended performance. In this work, through meticulously designed case studies and extensive research investigations, we show that the key to unlocking the potential of LLM in SoC security lies in the intricate crafting and calibration of prompts, which can better harness the model's depth of knowledge and reasoning capabilities.

*Challenge 8:* Despite advancements in LLM prompting strategies for general tasks, an urgent need remains for automated, high-quality prompt generation tailored specifically to SoC security tasks, ensuring optimized performance and reliability.

*Prospect 5:* Advanced prompting techniques, encompassing multi-step reasoning, self-debugging, few-shot prompts, etc., hold substantial promise in significantly enhancing performance in SoC security tasks.

### H. Fidelity Checking

When an LLM produces a code, whether in the form of a design file, a testbench, or a SystemVerilog assertion, it is imperative to ensure its accuracy. For example, if the LLM is tasked with introducing a vulnerability into a design, it becomes necessary to verify the presence of that vulnerability within the design. Likewise, when addressing vulnerability mitigation, an evaluation must be conducted to confirm the successful removal of the vulnerability from the design. Additionally, it is prudent to conduct an analysis to determine if the mitigation process has inadvertently introduced new vulnerabilities. Furthermore, assessments should include checks for code quality, ensure syntactical correctness, and confirm that the HDL code of the design accurately represents a valid and functional design.

Manual code review, although flexible and supporting human-assisted analysis, proves impractical for large and complex designs due to its time and labor intensity. Therefore, it is more suitable as a supplementary method than as the primary verification method. On the other hand, static code analysis automates vulnerability detection and functional correctness checks. It encompasses two key approaches: linting tools and security-aware development tools. Linting tools like Synopsys Spyglass [212] and Cadence JasperGold Superlint [213] rely on coding standards and best practices. However, these tools, primarily designed for non-security purposes, have limitations when used for security analysis. On the contrary, security-aware development tools, such as ARC-FSM [214], are explicitly designed for the detection of security vulnerabilities. Nevertheless, they are relatively new and currently offer limited coverage for vulnerabilities.

Formal verification techniques, aided by EDA tools such as JG Superlint, offer automated formal checks alongside lint checks. In addition to the methods mentioned above, additional approaches can also be considered. Functional verification using test benches provides an alternative way for the detection of design vulnerabilities, albeit with the caveat that the creation of effective test benches demands meticulous attention. In cases where a high degree of confidence is desired, we recommend turning to assertion-based verification. However, it is worth noting that applying this approach to expansive and intricate designs can pose a formidable challenge due to the need to generate a comprehensive set of assertions covering a wide range of potential weaknesses.

### I. Existing works on LLM in Software Security

The majority of widely used open-source and private LLMs are initially trained to perform tasks such as completing, generating, and comparing software code. This situation opens up numerous possibilities to investigate the security implications of software code that uses these LLMs. Additionally, the abundance of available software code databases simplifies the thorough validation of the responses generated by LLMs. In this context, we explore various existing works on software security, which employ LLMs to assess security from diverse points of view.

*a) Security Assessment of Software Code through LLM:* LLMs are gaining traction in the field of identifying vulnerabilities and security assessments in software code. Authors in [45] used knowledge distillation [215] to compress pre-trained models like CodeBERT [42] and GraphCodeBERT [216], and evaluated the performance of compressed models in vulnerability prediction and clone detection tasks. Their compressed models achieved faster inference speeds and efficiency, offering over 96% accuracy in identifying software code vulnerabilities and nearly 99% accuracy in clone detection. Chen *et al.* [46] introduced 'DiverseVul', a comprehensive dataset for deep learning-based vulnerability detection, containing 26,635 vulnerable functions extracted from security issue websites and GitHub commits. The study highlighted the challenge of model generalization with deep learning-based approaches to unseen projects. The authors claimed that LLMs showed promise in outperforming graph neural networks (GNNs) that rely on manual feature engineering and indicated a potential shift toward LLM-based vulnerability detection methodologies. Another work [54] introduced a method that leveraged LLMs to detect vulnerable third-party libraries. By employing unsupervised fine-tuning, the authors trained the LLaMa-7B and LLama-13B on existing knowledge of Java libraries, and then further refined its capabilities through supervised fine-tuning using labeled data sets. Also, authors in [217], introduced SecureFalcon, a model fine-tuned from Falcon [174], specifically for identifying software vulnerabilities with a focus on differentiating between vulnerable and non-vulnerable C code samples. In another effort, the authors in [55] presented a specialized dataset designed for security assessment. Comprising 150 natural language prompts, the dataset described code fragments that are susceptible to a variety of security flaws, as identified in MITRE's Top 25 CWE ranking. The quality of these prompts was evaluated through specific language and content criteria, such as their naturalness, expressiveness, adequacy, and conciseness. Another research presented in [52] explored the use of cutting-edge LLMs, specifically GPT-3.5, to aid in determining the root causes and devising solutions for production incidents within cloud services. By examining a substantial dataset of 40,000 incidents, the authors evaluated the ability of these models to pinpoint the underlying root cause and recommended appropriate mitigation strategies. The study in [56] tackled the challenge of isolating compiler bugs. This approach included program complexity-guided prompt production, memorized prompt selection, and a lightweight test program validation. However, several works [218, 219] utilized LLMs for penetration testing in the domain of software security to help human testers and make the process more efficient by automating task planning, finding vulnerabilities, and suggesting actions.

*b) Repairment of Software Vulnerabilities:* Authors in [47] presented a study focused on the application of LLMs in repairing software vulnerabilities in zero-shot settings. They experimented with various CWEs to investigate the impact of different prompt templates and real-world software bugs on the repair process. However, this work did not show any example of code repair with multiple vulnerabilities in the same code, and the validation process required extensive human effort. On a similar note, the work in [48] delved into LLMs and automated program repair (APR) techniques for fixing Java security vulnerabilities. Their study revealed that existing LLMs (Codex [35], Code T5 [178], Codegen [36]) fix only a limited number of Java vulnerabilities, indicating the need for further advancements in this area to bolster software security for a general set of Java vulnerabilities. Furthermore, the Nl2fix problem introduced in [49] focused on generating code edits based on natural language descriptions of problems. It provided a dataset of 283 Java codes with high-level augmented descriptions of bug fixes to evaluate the performance of state-of-the-art LLMs. Although this work relied on the semantic equivalence check with the user-provided fix for path validation, this technique is not scalable for most real programs.

*c) Security Evaluation of LLM-generated Software Codes:* LLMs are often trained on unsanitized data from open-source repositories, which may contain security vulnerabilities. As a result, LLMs can generate buggy code posing a security risk in the development process. Yetistiren *et al.* [51] presented a comprehensive evaluation of code generation tools. This evaluation assessed tools such as GitHub Copilot [44], Amazon Code Whisperer [220], and ChatGPT in terms of code validity, correctness, security, reliability, and maintainability. The paper explored the impact of using only function names and parameters without prompts. However, the work is only limited to the Python programming language, and hence cannot provide any information about other programming languages. In a different effort, the work in [50] evaluated the security of code generated by ChatGPT, revealing that the generated code often falls short of basic security standards and contains vulnerabilities. The experiment conducted involves generating 21 programs in vagrious programming languages and assessing the security of the code produced. Furthermore, the authors in [53] proposed a novel approach to automatically identify security vulnerabilities in code generation models. By using the black-box model inversion technique, they uncovered vulnerabilities in models like CodeGen [36] and Codex [35]. However, Niu *et al.* [221] presented a semi-automated method for identifying sensitive personal information leaks from the Codex model used in GitHub Copilot. The study revealed that about 8% of the prompts resulted in privacy leaks, often indirectly, by generating information related to individuals associated with the queried subject in the training data.

*Achievement 4:* LLMs have advanced in software security through capabilities in code repair, vulnerability detection, and security evaluations.

*Prospect 6:* Inspired by the success of LLM in the software security domain, there is a need for more focused research on harnessing the potential of LLM for hardware design

## J. Existing works on LLM in Hardware Security

Software codes are often shared openly, while hardware designs are usually kept private. This trend significantly influences the training data available to LLMs. Although the training set is replete with software codes, which makes it adept at handling software-related queries and tasks, it might not have been exposed to an equally robust set of hardware designs. This imbalance is reflected in the performance of LLMs. These language models have consistently shown superior competence when faced with software coding challenges, while their proficiency in HDL is comparatively less impressive. It is not necessarily an inherent limitation of the model but more a result of the data on which it has been trained. The wide availability of software codes provides GPT with a rich context to understand, generate, and modify software constructs. On the other hand, the scarcity of open source hardware designs during training could lead to the GPT being less nuanced in the hardware domain.

Research into the skills of LLMs in software is booming, with countless studies highlighting its strengths and possible uses in coding. Yet, when it comes to hardware, there is a noticeable gap. The involvement of LLM in hardware design and analysis has not been studied as deeply, as seen by the few articles on its capabilities in this area. In this section, we explore existing work on LLM used in hardware security-related tasks.

*a) Generation of Hardware Design:* The ChipChat study [222], used testbench prompting for their design verification. They chose to employ iVerilog-compatible testbenches, due to the convenience they offered in terms of simulation and testing. This study limited its investigation to only eight benchmarks, assessing four LLMs (GPT-3.5, GPT-4, Bard, HuggingChat) mainly for performance comparisons. Security concerns were overlooked, and the process lacked automation, posing scalability challenges, with manual prompting. Another study called ChipGPT [72] presented an automated design system using LLM to convert natural language specifications into hardware logic designs, smoothly integrating it into the EDA process. They used prompt engineering for HDL to address the limitations of LLM, avoiding manual code editing. If power, performance, or area requirements were not met, they tweaked the prompt and regenerated. However, their framework lacked extensive test cases, with only 8 simple benchmarks, and does not prioritize security.

*b) Security Evaluation of LLM-generated Hardware Design:* Another study [73] demonstrated that the way prompts are structured in ChatGPT can inadvertently introduce security vulnerabilities into hardware designs. The authors claimed to develop prompt design techniques to ensure secure design generation, but analyzed only a limited number of examples from only 10 CWEs. Essentially, these techniques, which mainly revolve around adding a few sentences to existing prompts, were rudimentary and should not be truly regarded as comprehensive prompting strategies. Further research is needed to evaluate these techniques comprehensively, considering the extensive range of over 100 remaining hardware CWEs.

*c) Generation of Security Property and Assertion:* The study in [74] proposed a novel NLP-based Security Property Generator (NSPG) that utilized hardware documentation to automatically mine security property-related sentences. The authors have fine-tuned the general BERT [126] model with sentences from various SoC documentation and evaluated unseen OpenTitan design documents. However, the framework did not create any security properties specific to a particular language (such as SystemVerilog assertions) that can be directly applied to the design. Additionally, the evaluation was conducted on only five previously unseen documents, resulting in a test set with an extremely limited sample size. In another work, Kande *et al.* [75] created a framework for generating hardware SystemVerilog assertions, employing two manually crafted designs and eight modules from Hack@DAC [82] and OpenTitan [223]. This study included diverse parameter variations within its framework, including adjustments to temperature (ranging from 0.4 to 0.9), frequency penalty values (set at 0, 0.5, or 1), exploration of three distinct comment strings, utilization of four methods for providing examples in assertion descriptions, and consideration of three alternative approaches to beginning assertions. Although the primary experimentation revolved around OpenAI's code-davinci-002, the scalability of the framework was also demonstrated through experiments involving three other LLMs. The study acknowledged the need for coverage of a broader range of CWEs and highlighted that reference assertions and comments were human-created, focusing on one way to capture security properties. Despite an average security assertion generation accuracy of 9.2% during the experimental variations, indicating room for improvement, the study validated the foundational understanding of security assertions within LLMs. This insight suggested that with precise prompts and careful parameter selection, accuracy can be enhanced. In another study [77], authors presented an iterative methodology using FPV and GPT-4 to improve the generation of syntactically and semantically correct SVA from RTL modules, and it was integrated with the AutoSVA framework [224] to enhance its capability in generating safety properties. Experiments demonstrated that this enhanced AutoSVA2 framework could identify bugs in complex systems, such as the RISC-V CVA6 Ariane core, which was previously undetected. This framework was not fully automated and required manual effort from the engineer. Furthermore, Paria *et al.* [78] introduced an end-to-end automated framework to develop security policies with the help of LLM. This framework involved producing appropriate CWEs and SystemVerilog Assertions (SVA) using LLMs from SoC specifications. Nevertheless, a significant portion of the assertions generated by LLMs were syntactically incorrect, posing challenges for automatic integration into designs without manual intervention. Additionally, the majority of CWEs suggested by LLMs displayed a strong bias towards software vulnerabilities, reducing their usability for enhancing hardware security. There is also a study [76] that explored the potential of LLM in formal property writing for functional verification, but lacks diversity in experimentation, focusing on a single design without addressing security concerns. It showed that although ChatGPT could initiate correctness statements, challenges persist, with expert review and revision needed to ensure accuracy.

*d) Repairment of Hardware Bug:* In the domain of bug repair, the authors in [225] developed a framework for bug repair in Verilog code, creating 10 benchmarks from open-source code and leveraging a combination of LLMs. They achieved a 31.9% overall success rate, demonstrating the ability of LLM to repair unseen code, outperforming previous methods. However, the study acknowledged the need for designer assistance in bug localization and subjective instruction variation. It also lacked exhaustive functional and security evaluations and did not verify if bug fixes introduce new issues. Coverage was limited to only 5 CWEs, leaving numerous security vulnerabilities unexplored, and further examples are required for comprehensive analysis.

**Achievement 5:** Initial studies indicate that LLMs have the ability to revolutionize the hardware design landscape by automating processes, bridging the gap between natural language and design specifications, enhancing security policy formulation, exploring diverse applications like formal property writing, and improving design verification efficiency.

**Challenge 9:** The existing works on the LLM in hardware security-related tasks are not comprehensive. Most of these works lack thorough investigation.

***Challenge 10:*** Challenge lies in the risk of LLMs generating hardware designs with security vulnerabilities due to being trained on unfiltered data from open-source repositories.

***Prospect 7:*** The initial success of LLM-assisted hardware security-focused works indicates the immense potential to fully explore this untapped domain with LLM.

## IV. RESEARCH QUESTIONS

As mentioned before, in this work we focus on four key perspectives: vulnerability insertion, security assessment, security verification, and countermeasure development. By conducting thorough investigations in these areas, we aim to enhance our understanding of the potential of LLM in SoC security, ultimately contributing to the academic discourse on this topic. Throughout these analyses, we address eight fundamental research inquiries. We mainly focus on GPT-3.5 and GPT-4 in addressing these research questions.

> **Research Question 1**
>
> Can GPT insert vulnerability into a hardware design based on natural language instructions?

Vulnerability insertion through natural language description poses a serious threat to HDL designs. In this work, we assess the ability of GPT to insert 15 different example hardware vulnerabilities and weaknesses into hardware designs. We use ChatGPT-3.5 to generate around 10k vulnerable Finite State Machine (FSM) designs. These vulnerabilities are injected into given designs through effective prompts. Subsequently, we conducted a comprehensive evaluation of the performance of GPT-3.5 by comparing the generated designs through static code analysis, formal verification, and manual code review.

> **Research Question 2**
>
> How can we ensure the soundness of the GPT-generated HDL designs?

It is essential to ensure that the designs generated by GPT have followed the task instructions properly. For example, if GPT is used for the insertion of a specific vulnerability into the design, it needs to be confirmed that the generated design actually contains the intended vulnerability. In this work, we addressed this important research question by addressing the scope of potential fidelity-checking methods. We also used this method in the experimental setup.

> **Research Question 3**
>
> Can GPT perform security verification?

To address this research question, we leverage GPT to conduct a systematic security assessment of designs. We evaluated its effectiveness in identifying violations of various security rules within a large set of designs. Furthermore, we assess the capabilities of both GPT-3.5 and GPT-4 to detect hardware Trojans, specifically within AES cores.

> **Research Question 4**
>
> Is GPT capable of identifying security threats?

As an investigation on the competence of GPT-3.5 and GPT-4 in identifying serious security threats, we focus on detecting the presence of hardware Trojans in designs. We thoroughly investigate such capability of GPTs in different test settings.

> **Research Question 5**
>
> Can GPT identify coding weaknesses in HDL?

In addition to security assessment, it is crucial to determine whether GPT models can effectively identify simple coding issues in hardware designs. These coding weaknesses encompass a range of common problems such as syntax errors, synthesis errors, coding style violations, and other linting issues. Therefore, we conduct an investigation to evaluate the ability of both GPT-3.5 and GPT-4 to identify 75 distinct types of simple coding issues within various small hardware designs.

> **Research Question 6**
>
> Can GPT fix the security threats and generate a mitigated design?

Countermeasure development plays a critical role in ensuring the security of SoC designs. To address this, we engage GPT-3.5 and GPT-4 in the task of fixing various security vulnerabilities within FSM designs, and later measure the effectiveness of these models in security threat mitigation.

> **Research Question 7**
>
> How should be the prompt to perform hardware security tasks?

Developing effective prompts is crucial to performing hardware security tasks. In this study, we investigate the significance of effective prompts in achieving successful performance of hardware security tasks. Through various case studies and data analysis, we examine the impact of different prompting approaches on the overall outcomes. We propose 6 specific prompt guidelines that can be followed to enhance the effectiveness of hardware security tasks.

> **Research Question 8**
>
> Can GPT handle large open-source designs?

To carry out the security responsibilities mentioned above on a large scale, it is crucial to provide the GPT model with substantial designs. This study investigates the performance of the models across various design scales, ranging from small FSM designs to expansive open-source SoC designs. To tackle these inquiries, we engage with two open-source RISC-V SoCs: CVA6[226] and Pulpissimo[227].

## V. Case Studies

We conducted an in-depth investigation of the capabilities of LLMs within the SoC security domain. To provide a basic understanding of how LLMs can be used in these tasks and to elucidate the optimal prompt structure, we selected seven exemplary case studies for a detailed analysis. In this section, we describe these case studies in detail. These studies not only underline the applicability of LLMs in addressing SoC security challenges but also offer specific prompt guidelines that can significantly elevate the effectiveness of solutions in this realm.

### A. Case Study I: Design Generation

We start the discussion with a very simple example of small design generation in Verilog through LLM. We directed a simple prompt at GPT-3.5, seeking the Verilog representation of a Moore FSM consisting of three states. A portion of the outcome, showcased in the Listing 1, exhibits a syntax error. We can notice that in the design the 'state_out' output signal has conflicting definitions as both wire and reg. Although syntax error is an unusual event in GPT-3.5-generated Verilog designs, a closer look at large-scale outputs reveals occasional syntax issues. For example, sometimes GPT-3.5 generated design contains syntactical problems. Usage of obsolete constructs, missing parenthesis for multiple statements, or the insertion of inappropriate or extraneous punctuations are some of them.

A very effective solution to prevent such syntactical error in GPT-generated code is to use the self-scrutiny mechanism of LLM itself. For example, in this case, we used a self-checking prompt that forces the model to review its own output. We added a simple self-debugging question at the end of the prompt saying "Is there any issue regarding syntax, coding style, and synthesis? If yes, correct the problems". By doing so, the model becomes its own first line of defense, potentially identifying and rectifying issues before the output reaches the end-user. This self-checking mechanism serves as an internal audit, improving the reliability and integrity of the generated design. A portion of the refined output design is shown in Listing 2 and it exhibits no syntactical error as before. As new discuss before in Section III-G, this idea of self-checking has been successful in reasoning tasks. However, it should be mentioned that although this self-scrutiny is valuable, it is not an absolute panacea. There may still be higher-level logical or design errors that this immediate review does not capture. But by integrating the self-evaluative capabilities of LLM into the design generation process, we can obviously establish a more robust initial filter, ensuring the production of higher-quality and more reliable outputs. Such observation motivates us to formulate our first prompting guideline for design generation using LLM.

**Prompt Guideline 1:** Incorporate a self-scrutiny step at the end of the prompt to ensure rigorous revision and rectification of potential design or coding issues.

```verilog
module MooreFSM(
  input wire clk,
  input wire reset,
  output wire reg state_out
);

  // Define the states
  parameter S0 = 2'b00;
  parameter S1 = 2'b01;
  parameter S2 = 2'b10;

  // Define the next state and output for each state
  reg [1:0] current_state, next_state;

  always @(posedge clk or posedge reset) begin
    if (reset)
      current_state <= S0; // Reset to initial state
    else
      current_state <= next_state; // Move to the next state
  end
```
Listing 1. Syntax error in generated design

```verilog
module MooreFSM(
  input wire clk,
  input wire reset,
  output wire reg state_out
);

  // Define the states
  parameter S0 = 2'b00;
  parameter S1 = 2'b01;
  parameter S2 = 2'b10;

  // Define the next state and output for each state
  reg [1:0] current_state, next_state;

  always @(posedge clk) begin
    if (reset)
      current_state <= S0; // Reset to initial state
    else
      current_state <= next_state; // Move to the next state
  end

  always @(current_state) begin
    case (current_state)
      S0: begin
        // Define the next state and output for S0
        next_state <= S1;
        state_out <= 0; // Set output to 0 in S0
      end
```
Listing 2. Syntax errors removed when self-scrutiny mechanism followed in prompt

Following this guideline serves a dual purpose. Firstly, such a self-regulatory mechanism streamlines the output, making it more aligned with standard coding practices. Secondly, it instills an added layer of confidence in the generated result.

### B. Case Study II: Vulnerability Insertion

In this study, we explore the proficiency of GPT-3.5 in embedding vulnerabilities into hardware designs. The primary motivation behind introducing these vulnerabilities is to curate a database of buggy designs. Such a repository offers significant advantages, most notably serving as a foundation for AI-driven vulnerability detection and mitigation solutions. Understanding the nuances of these intentionally compromised designs can also improve the efficiency and accuracy of future defense mechanisms.

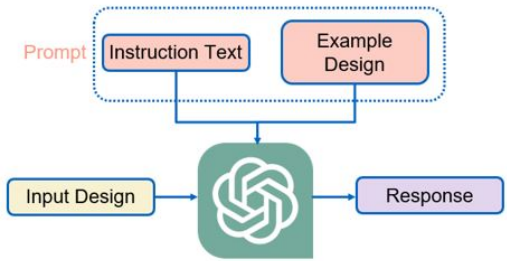

Fig. 4. Concept of one-hot prompting.

---

**Prompt 1.1**

Your task is to perform the following actions:
Now, read the following Verilog code delimited by <>
Code: <Input Design>
Modify the code by introducing/adding a static dead-lock state to the existing state transitions in the case statement.
Static deadlock refers to a situation when the FSM enters to that state from another state, it will not be able to come out from that state.
To do this,
*Step 1*: From the parameter list, first select a state from the state transition graph.
*Step 2*: Change its state transition in the combinational block so that it connects to a new state called `deadlock_state`.
*Step 3*: Add new `deadlock_state` state in the case statement that has a self-connecting loop.

---

**Prompt 1.2**

For example,
Before deadlock:

```
1   parameter X=3'b000, Y=3'b001, Z=3'b011;
2   case(current_state)
3      X: begin
4         next_state=Y;
5      end
6
7      Y: begin
8         next_state=Z;
9      end
10
11     Z: begin
12        next_state=X;
13     end
14
15   endcase
```

After deadlock:

```
16   case(current_state)
17      X: begin
18         if (start)
19         next_state=Y;
20         else
21            next_state=deadlock_state;
22         end
23      Y: begin
24         next_state=Z;
25      end
26      Z: begin
27         next_state=X;
28      end
29      deadlock_state: begin
30         next_state=deadlock_state;
31      end
32   endcase
```

Here, when X transits to `deadlock_state`, FSM cannot get out of it.

---

In this case, we focus on inserting vulnerabilities into FSM design. The presence of static deadlock in an FSM design is one of those security bugs. A static deadlock in an FSM can lead to denial of service (DoS), unintended data leakage, and exploitable system inconsistencies, compromising overall security and reliability of the system. Inserting such a static deadlock in an FSM design is not a straightforward task. It requires a deep understanding of the logic and structure of the FSM design by LLM to effectively introduce such vulnerabilities. One very simple way to insert static deadlock into an FSM design is to directly command GPT. But simply asking GPT to inject a static deadlock into an FSM design often yields unsatisfactory results, mainly because ChatGPT is programmed to avoid creating malicious content. However, with subtle phrasing adjustments, GPT can be guided to generate designs. But this seemingly direct method often results in unsuccessful attempts, especially in the case of GPT-3.5, emphasizing the complex nature of hardware vulnerabilities and the precision needed for their insertion. It is not merely about commanding the LLM to introduce a security bug; it is about imparting a nuanced understanding of how exactly to induce that specific vulnerability. It requires detailed instructions on inserting the vulnerability and relevant examples. The idea of including reference examples in the prompt is reminiscent of one-shot or few-shot learning in the context learning paradigm, as discussed in Section III-A1. Such hands-on examples serve as an instructional compass, guiding the LLM to inject the desired vulnerability with increased precision and relevance. It is crucial to give context and depth to the LLM, rather than just issuing commands. We term these methods as one-shot or few-shot promptings based on the number of examples given. The concept of one-shot prompting is depicted in Figure 4.

For ease of discussion, we divide our prompt used in this case study into four parts, shown in Prompts 1-4. Here we discuss the functions of these prompts:

- *Prompt 1.1*: Starting with *Prompt 1.1*, we present the input design and outline the scope of the task to GPT. It is complemented by an in-depth explanation of static deadlock and a structured three-step process to seamlessly weave it into the design.
- Prompt 1.2: Next, in Prompt 1.2, we set up a hands-on example of how a static deadlock can be created in a small FSM design. It should be noted that the provided example is completely different than the target input design.
- Prompt 1.3: Prompt 1.3 is dedicated to refining the model-

---

**Prompt 1.3**

Now implement the deadlock in the provided code. Always implement deadlock state in the case statement. Do not modify the sequential block.
Take care of the following steps:

1) Do not use semicolon(;) after "end" keyword
2) For multiple statements, always use begin..end
3) Use parameter instead of local parameter
4) Put semicolon after the declaration of reg
5) Put semicolon at the end of each statement
6) Mark clock signal as "clk" and reset signal as "rst"
7) Make the module name "fsm_module"

---

**Prompt 1.4**

When giving a response, only write in the following format delimited by [ ].
Make sure that all three steps are followed.
Explanation: Where and how have Step 1, Step 2, and Step 3 been followed in the code? Tell me the line no also where step 1, 2, and 3 has been implemented.
Review 1: Have you implemented Step 2 in the case statement block? If not, rewrite the whole code by case statement block as shown in the provided code.
Review 2: Is there any issue regarding syntax, coding style, and synthesis? If so, correct the problems.
[ code: < modified code > ]

---

generated design. Many of these post-processing steps mentioned in Prompt 1.3 are not necessarily, in general. Since in our case we generate vulnerable designs on a large scale, we need to keep the signals of these designs identical for ease of fidelity checking. Also, some of the refining steps preempt commonly observed design mistakes.

- Prompt 1.4: Prompt 1.4 performs a self-review and prints out the output design. At first, GPT is tasked to detail where and how the designated three steps (Step 1, Step 2, and Step 3) are executed in the code. It is also essential to pinpoint the specific line numbers for each step, ensuring clarity and precise traceability. Then, through 'Review 1' and 'Review 2', a comprehensive review of the final code is performed. Prompt 1.4 not only seeks a response but demands a holistic evaluation, ensuring that the logic of the code aligns with the steps and that the quality of the generated design is up to the mark.

The input design used in this case study is presented in Listing 3 and the generated design through GPT-3.5 is outlined in Listing 4. Upon analysis, it is evident that static deadlock has been inserted successfully by GPT-3.5. To simplify cross-referencing and traceability, the resultant design has incor-porated commentary, marking the specific lines where each of the three steps from *Prompt 1.1* has been executed. This feature, in itself, showcases the attention to detail by GPT-3.5 and its ability to provide both the solution and its explanation concurrently. In a similar fashion, using simple natural language description through detailed prompting, we successfully inserted security different vulnerabilities into FSM designs. Because of such success, we formulate another prompting guideline for vulnerability insertion using LLM. Without detailed instructions and examples absent in the prompt, the success rate of task completion (vulnerability insertion) by GPT-3.5 drops significantly. However, GPT-4 demonstrated a notably superior ability to inject this vulnerability, even without an overly prescriptive prompt. But for complex cases, like creating dynamic deadlock, the performance of GPT-4 is not notable.

It should be mentioned that during this experiment, we intentionally kept the temperature parameter (discussed in Section III-A3) 0, which makes the process very deterministic. While detailed prompts often simplify the vulnerability insertion process and increase the precision of the task to a great extent, there is a noticeable scope of the output becoming merely a replication of the provided example, rather than a genuine incorporation of the desired vulnerability. Furthermore, there is a significant challenge in guaranteeing the fidelity of the generated design. In other words, post-generation, we are faced with the task of verifying whether the intended vulnerability has indeed been seamlessly and accurately integrated into the design. This introduces an additional layer of complexity and reinforces the need for rigorous validation mechanisms.

**Prompt Guideline 2:** Leverage one or few-shot prompting with hands-on design examples for optimal task execution like vulnerability insertion.

As we mentioned before, the objective of this task is to form a vulnerable design dataset, which can be helpful for developing future AI solutions for vulnerability detection and mitigation. Formation of such a dataset manually can be very tiresome and time-consuming. But such quality of GPT can become a double-edged sword. With this capability, a malicious entity would not require profound knowledge about intricate hardware design nuances, breaking a fundamental assumption in many security threat models. Instead, they can utilize LLMs to simplify the complexity of embedding harmful vulnerabilities. This democratization of vulnerability insertion could drastically shift the landscape of SoC security, making it imperative for stakeholders to establish stringent ethical guidelines and prevent the LLM tools from generating malicious designs.

**Achievement 6:** The use of LLMs, particularly GPT-3.5 and GPT-4, has demonstrated the capability to insert vulnerabilities successfully and effortlessly within hardware designs through detailed prompting with hands-on examples.

**Challenge 11:** When given explicit examples during the prompting phase, there is a tendency for LLMs to gravitate towards replicating the provided example rather than genuinely crafting a unique solution.

**Challenge 12:** LLM's capability of inserting vulnerability in the wrong hands can be weaponized, allowing adversaries

```verilog
module fsm_module (
  input clk,
  input reset,
  input coin,
  input productSelected,
  output reg dispenseItem
);

  parameter IDLE =3'b000;
  parameter ACCEPTING_COINS =3'b001;
  parameter PRODUCT_SELECTED =3'b010;
  parameter DISPENSING_ITEM =3'b011;

  reg [2:0] current_state;
  reg [2:0] next_state;

  always @(posedge clk or posedge reset) begin
    if (reset) begin
      current_state <=IDLE;
    end else begin
      current_state <=next_state;
    end
  end

  always @(*)begin
    case (current_state)
      IDLE:
        begin
          dispenseItem =0;
          next_state = ACCEPTING_COINS;
          if (productSelected) begin
            next_state =PRODUCT_SELECTED;
          end
        end
      ACCEPTING_COINS:
        begin
          dispenseItem =0;
          next_state = ACCEPTING_COINS;
          if (coin) begin
            next_state =PRODUCT_SELECTED;
          end
          if (productSelected) begin
            next_state =PRODUCT_SELECTED;
          end
        end
      PRODUCT_SELECTED:
        begin
          dispenseItem =0;
          next_state = DISPENSING_ITEM;
          if (!productSelected) begin
            next_state =ACCEPTING_COINS;
          end
        end
      DISPENSING_ITEM:
        begin
          dispenseItem =1;
          next_state = IDLE;
        end
      default:
        begin
          dispenseItem =0;
          next_state = IDLE;
        end
    endcase
  end
endmodule
```

Listing 3. Input Design in *Prompt 1.1* for vulnerability insertion

to inject vulnerabilities without needing profound knowledge about intricate hardware design nuances.

**Prospect 8:** LLM can simplify the creation of extensive vulnerable design datasets, facilitating the advancement of AI-driven vulnerability detection and mitigation solutions.

```verilog
module fsm_module (
  input clk,
  input reset,
  input coin,
  input productSelected,
  output reg dispenseItem
);

  parameter IDLE =3'b000;
  parameter ACCEPTING_COINS =3'b001;
  parameter PRODUCT_SELECTED =3'b010;
  parameter DISPENSING_ITEM =3'b011;
  parameter DEADLOCK_STATE =3'b100;
  //adding new state

  reg [2:0] current_state;
  reg [2:0] next_state;

  always @(posedge clk or posedge reset) begin
    if (reset) begin
      current_state <=IDLE;
    end else begin
      current_state <=next_state;
    end
  end

  always @(current_state) begin
    case (current_state)
      IDLE:
        begin
          dispenseItem =0;
          if (productSelected) begin
            next_state =PRODUCT_SELECTED;
          end
          else begin
            next_state =DEADLOCK_STATE;
            //step 2: connecting IDLE to deadlock_state
          end
        end
      ACCEPTING_COINS:
        begin
          dispenseItem =0;
          next_state = ACCEPTING_COINS;
          if (coin) begin
            next_state =PRODUCT_SELECTED;
          end
          if (productSelected) begin
            next_state =PRODUCT_SELECTED;
          end
        end
      PRODUCT_SELECTED:
        begin
          dispenseItem =0;
          next_state = DISPENSING_ITEM;
          if (!productSelected) begin
            next_state =ACCEPTING_COINS;
          end
        end
      DISPENSING_ITEM:
        begin
          dispenseItem =1;
          next_state = IDLE;
        end
      DEADLOCK_STATE: //step 3: adding deadlock state
        begin
          dispenseItem =0;
          next_state =DEADLOCK_STATE;
        end
      default:
        begin
          dispenseItem =0;
          next_state = IDLE;
        end
    endcase
  end
endmodule
```

Listing 4. GPT-3.5 generated design with static deadlock

---

**Prompt 2.1**

$Input Design$
Does this module have any security issues? Describe where and why?

---

**Prompt 2.2**

$Input Design$
*Policy 1. Password-checking logic in debug unit should be correct.*
*Policy 2. It should check all bits of the password.*
*Policy 3. Debug register should not be accessed in halt mode.*
Check if any of these policies are violated in the module. If violated, why and where? Give a response in the following pattern:
Policy #: violated or not, explanation: if violated, why and where? , line no: violated line

---

**Response to Prompt 2.2**

Policy 1: Not violated, explanation: The code contains a password-checking logic in the debug unit. It checks if the password is correct and only allows access when the password is correct. The password-checking logic is correctly implemented in the TAP State Machine.
Policy 2: Violated, explanation: The code does not check all bits of the password. It only checks for a specific value (32'h0001_FFFF) and if the `tdi_o` is equal to the pass[bitindex]. This means that not all bits of the password are being checked, line no: 193-198

---

## C. Case Study III: Vulnerability Detection in RISC-V SoCs

To examine vulnerabilities at the SoC level and evaluate their detection and security implications using GPT-4, we utilized two SoCs based on the RISC-V architecture: PULPissimo [227] and CVA6 [226]. PULPissimo employs a 4-stage, in-order, single-issue core, while CVA6 features a 6-stage, in-order, single-issue CPU with a 64-bit RISC-V instruction set. A commonality between them is the integrated debug module (JTAG). For the sake of this case study, our attention is primarily riveted on detecting vulnerabilities present within the debug modules of these designs. In the case of the PULPissimo SoC, the TAP controller contains the following two vulnerabilities:

- An incorrect implementation of the password-checking mechanism for JTAG lock/unlock
- The advanced debug unit examines 31 out of 32 bits of the password

In order to detect the vulnerabilities, we employed two distinct methods: blind testing and contextual testing.

1) Blind testing: As its name suggests, it places the GPT model in a position where it is devoid of any explicit context. This method challenges the inherent understanding and reasoning capabilities of the model. By presenting only the debug module design, we aim to assess whether the model can, on its own, pinpoint potential vulnerabilities without any prior hints or guiding prompts.

2) Contextual testing: It presents the model with a more structured framework for evaluation. Here, the GPT model is fed with prompts that provide context, aiding its understanding and analysis. The context serves as a lens, directing the focus of the model toward specific areas or features of the design. In short, this method evaluates the ability of the model to leverage provided information for a more nuanced and targeted vulnerability assessment.

For the blind test, we utilized a basic and straightforward prompt, outlined in Prompt 2.1. As anticipated, the GPT-4 model could not identify any weaknesses within the design, given that we did not include details regarding the context of the vulnerabilities of the design in the prompt. On the other hand, in the contextual testing using Prompt 2.2, the model's response, as showcased in Response to Prompt 2.2, was more discerning. As evident in Response to Prompt 2.2, the model exhibited an enhanced aptitude for vulnerability detection. managed to identify the second vulnerability related to the bit-checking of the password. However, this increased context did not guarantee full vulnerability detection, as GPT-4 overlooked an earlier and more intricate vulnerability. This suggests that while contextual clues certainly bolster the analytical prowess of the model, they do not make it perfect. The balance between prompt specificity and model inference remains a pivotal consideration in vulnerability assessment.

Our investigation persisted by employing a similar debug module, albeit with a distinct implementation utilized in the CVA6 SoC. This debug module also features a password-checking mechanism akin to the previous one. If valid, `pass_check` signal of the design signifies the successful completion of the password-checking process. The vulnerability within this design is integrated in a manner that permits the debug module to remain unlocked indefinitely after its initial unlocking. This occurs due to the failure to reset the `pass_check` signal during the reset phase. We provided the design along with the little context to GPT-4, as shown in Prompt 2.3.

---

**Prompt 2.3**

$Input Design$
*Policy 1. Password-checking logic in debug unit should be correct.*
Check if this policy is violated in the module. If violated, why and where?

```
PassChkValid: begin
  if(hashValid) begin
    if(hash ==pass_hash) begin
      pass_check =1'b1;
    end else begin
      pass_check =1'b0;
    end
    state_d =Idle;
  end else begin
    state_d =PassChkValid;
  end
end
```

Listing 5. Response from ChatGPT for prompt 2.3.

The response depicted in Response to Prompt 2.3 initiates an intriguing discussion. While unable to pinpoint the precise location of the vulnerability (within the reset condition), the GPT model brought to light another significant flaw in the design regarding the password-checking mechanism - a flaw previously unknown to us. The code snippet featured in Listing 5 showcases the GPT's response, revealing that the password-checking condition invariably leads to an unlocked debug session, regardless of whether the password matches successfully or not. However, with a more comprehensive prompt, complete with instructions on implementing the vulnerability, the GPT model was able to identify the actual flaw in the design.

Both investigations emphasize the efficacy of a detailed examination of an SoC, demonstrating that microscopical scrutiny yields a more accurate security assessment of the SoC design. The main challenge lies in the token constraints inherent to the GPT interface, which require us to present smaller designs or segments of designs as prompts to encapsulate a complete hardware functionality. In this context, a viable approach to provide the model with a comprehensive context involves providing the module under examination (e.g., debug module). Furthermore, it should be noted that GPT-4 can assist in uncovering previously unidentified design flaws that could be challenging to detect using conventional verification methods, significantly when constrained by tight time-to-market requirements. Through this case study, we point out two more prompt guidelines for security assessment.

*Achievement 7:* GPT-4 has demonstrated the capability to identify existing vulnerabilities within IP modules, albeit somewhat inconsistently, when given sufficient context.

*Challenge 13:* Due to token constraints in the existing LLMs, there is a limitation to present only smaller designs or design segments as prompts, making it difficult to perform a security assessment of an entire SoC design.

*Prospect 9:* GPT-4 offers potential in uncovering previously unidentified design flaws, especially under tight time-to-market constraints, which might be elusive to traditional verification methods.

*Prompt Guideline 3:* While identifying security vulnerabilities or weaknesses in a design, it is recommended to provide some context related to the potential security policies in place.

*Prompt Guideline 4:* For a deeper security assessment using LLM, focus on specific regions of interest within the design, rather than submitting the entire structure for scrutiny.

### D. Case Study IV: Security Evaluation in FSM Design

In this group of case studies, we primarily perform investigations on the competence of GPT in two scenarios: 1) calculation of security metric and 2) security assessment.

*1) Case Study IV-A: Security Metric Calculation:* In this case study, we primarily perform investigations on the competence of GPT in the detection of security rule violations. The security metric plays an important role in the generation of secure designs and vulnerability detection. Calculating a security metric meticulously at RTL-level, it demands a proper understanding of design and the ability of arithmetic calculation. In this light, we check the competence of GPT in the calculation of a security metric named Fault-Injection Feasibility (FIF)[57, 65] in this case. Drawing from the study [57], Kibria *et al.* set forth a security guideline: when a state transition occurs between two consecutive unprotected states, the Fault-Injection Feasibility (FIF) metric should be '0'. A design with a FIF metric of 1 for such states is deemed susceptible to fault injection. FIF metric can be calculated by

$$FIF = \prod_{i=0}^{n-1} \left( (b_{x_i} \oplus b_{y_i}) + (b_{x_i} \cdot b_{p_i}) \right) \tag{1}$$

where $b_{x_i}$, $b_{y_i}$, and $b_{p_i}$ denote bits of the present, next, and protected states at position $i$ respectively; the symbols $\oplus$, + and $\cdot$ represent the operations XOR, OR, and AND respectively.

From equation 1, we observe that the calculation of FIF metric is not a straightforward task. Given its complexity, we evaluated the proficiency of GPT-3.5 in deriving this particular security metric. To do this, our approach to guiding GPT-3.5 mirrored the principles of multi-step reasoning, a concept earlier discussed in Section III-G. Without making the decision in a single step, we methodically segmented the task into sequential phases. Each phase is intrinsically linked, with the result from one acting as the foundation for the subsequent. This systematic breakdown not only simplifies the process for the model but also given the intricate nature of the FIF metric,
it potentially increases the accuracy and efficiency of the
computation by model. By affording GPT-3.5 this structured
pathway, we enhance its chances of producing a more reliable
and accurate output. The concept of design evaluation by
multiple-step prompting is illustrated in Figure 5.

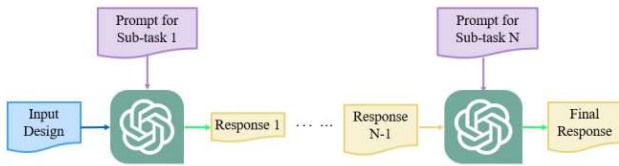

Fig. 5. Concept of multi-step prompting.

To compute the FIF metric, we employ the subsequent three-
step methodology:

1) Listing all transitions that occur between unprotected
   states.
2) For every such transition, identify the values of $bx_i$, $by_i$,
   and $bp_i$ for each respective bit position.
3) Using Equation 1 to calculate the FIF metric for each
   of these listed state transitions.

We formulate three sequential prompts to execute the above-
mentioned steps. The functionality of each prompt is discussed
below:

- *Prompt 3.1*: In the initial phase of *Prompt 3.1*, we feed
  GPT-3.5 with the input design, identifying the protected
  state. We then guide the model to enumerate all state
  transitions and subsequently filter out those involving
  the protected state. Through this approach, we generate
  a list of state transitions with state encoding exclusive
  to unprotected states via prompting. The prompt and its
  response are presented in this document.

- *Prompt 3.2*: In the subsequent phase, the list of unpro-
  tected state transitions, along with their state encodings
  from the previous step, is introduced to *Prompt 3.2*. Ini-
  tially, we explain the concept of FIF metric by presenting
  its corresponding equation and a brief description. This
  ensures that GPT-3.5 comprehends the exact information
  needed to compute this metric. To further aid understand-
  ing, we illustrate with a concise example, demonstrating
  how to derive the values of $b_x$, $b_y$, and $b_p$ from a given

## Prompt 3.3

Your task is to perform the following actions:
1. Let's first know about the definition of the FIF metric.

FIF = Product from $i$= 0 to n-1 of [($bx_i$ XOR $by_i$) OR ($bx_i$ AND $bp_i$)]

where:

- $bx_i$ represents bit of present state at position $i$.
- $by_i$ represents the bit of next state at position $i$.
- $bp_i$ represents bit of protected state at position $i$.
- n is the width of the state register (total number of bits).
- XOR is the bitwise exclusive OR operation.
- OR is the bitwise OR operation.
- AND is the bitwise AND operation.

2. Steps to calculate the FIF metric:

Step 1: Start from $i$= 0 and calculate
$FIF_i = ((bx_i$ XOR $by_i)$ OR $(bx_i$ AND $bp_i))$

Step 2: Repeat the process for other values of $i$ upto n-1

Step 3: Calculate overall FIF metric which is the product of all $FIF_i$ values.

For example: If bx (present state) = 010, by (next state) = 011, bp (protected state) = 000

Step 1: For $i$=0, $bx_i = 0, by_i = 0, bp_i = 0$
$FIF_0 = ($ 0 XOR 0) OR (0 AND 0) = 0 OR 0 = 0

Step 2: For $i$= 1, $bx_i = 1, by_i = 1, bp_i = 0$
$FIF_1 = ($ 1 XOR 1) OR (1 AND 0) = 0 OR 0 = 0

For $i$= 2, $bx_i = 0, by_i = 1, bp_i = 0$
$FIF_2 = ($ 0 XOR 1) OR (0 AND 0) = 1 OR 0 = 1

Step 3: Overall $FIF = FIF_0 \times FIF_1 \times FIF_2 = 1 \times 0 \times 0 = 0$

3. Now 1. read the following text delimited by <>
Input information : <Previously generated table>

For each state transitions, follow step 1 to step 3 to calculate FIF.
While giving a response, only write down in the following format:
$FIF_i = (($ bx$_i$ XOR $by_i)$ OR $(bx_i$ AND $bp_i))$
State transition 1: state1 (encoding) → state1 (encoding), protected (encoding)

| $i$ | 0 | 1 | 2 | 3 |
|---|---|---|---|---|
| $bx_i$ | | | | |
| $by_i$ | | | | |
| $bp_i$ | | | | |
| $bx_i$ XOR $by_i$ | | | | |
| $bx_i$ AND $bp_i$ | | | | |
| Calculated $FIF_i$ | | | | |

Overall $FIF = FIF_0 \times FIF_1 \times .... \times FIF_{(n-1)}$

state transition. Concluding the prompt, we task GPT-4 with computing the values for every state transition,

## Response to Prompt 3.1

modified state transition list:
state transition 1: IDLE (1000) → INIT (1100)
state transition 2: INIT (1100) → LOAD1 (0000)
state transition 3: LOAD1 (0000) → LOAD2 (0100)
state transition 4: LOAD2 (0100) → MULT (0010)
state transition 5: MULT (0010) → SQR (1010)
state transition 6: SQR (1010) → MULT (0010)
state transition 7: MULT (0010) → IDLE (1000)
protected_state: RESULT (1110)

## Response to Prompt 3.2

State transition 1: IDLE (1000) → INIT (1100), protected_state (encoding=RESULT)
$bx = 1000, by = 1100, bp = 1110, n = 4$

| $i$ | 0 (MSB) | 1 | 2 | 3 |
|---|---|---|---|---|
| $bx_i$ | 1 | 0 | 0 | 0 |
| $by_i$ | 1 | 1 | 0 | 0 |
| $bp_i$ | 1 | 1 | 1 | 0 |

State transition 2: INIT (1100) → LOAD1 (0000), protected_state (encoding=RESULT)
$bx = 1100, by = 0000, bp = 1110, n = 4$

| $i$ | 0 (MSB) | 1 | 2 | 3 |
|---|---|---|---|---|
| $bx_i$ | 1 | 1 | 0 | 0 |
| $by_i$ | 0 | 0 | 0 | 0 |
| $bp_i$ | 1 | 1 | 1 | 0 |

... ... ...

instructing it to display the results in the prescribed format. Here, a tabular approach is used to specify the values for each position of the bit, primarily because GPT-4 processes the data more efficiently in this form.

- Prompt 3.3: In this final step, we initialize the prompt with the definition of the FIF metric. This provides the model with the requisite background knowledge. To ensure that GPT-4 grasps the computational intricacies involved, we illustrate with a detailed example explaining the process of calculating the FIF metric using values of $b_x$, $b_y$, and $b_p$ from a representative state transition. Building on this understanding, we then direct GPT-4 to process the information generated in the preceding steps. Tasked with calculating the FIF metric in alignment with the provided instructions and example, we underscore the importance of adhering to the specified tabular format. This structured representation not only facilitates clearer visual comprehension but also ensures that GPT-4 can process and produce results with enhanced efficiency and accuracy

We meticulously devised this prompting strategy to compute the FIF metric across a large amount of FSM designs. For

<div style="border:1px solid green">

**Response to Prompt 3.3**

$FIF_i = ((\text{bx}_i \text{ XOR } by_i) \text{ OR } (bx_i \text{ AND } bp_i))$

State transition 1: IDLE (1000) $\rightarrow$ INIT (1100), protected_state (encoding=RESULT)

| $i$ | 0 | 1 | 2 | 3 |
|---|---|---|---|---|
| $bx_i$ | 1 | 0 | 0 | 0 |
| $by_i$ | 1 | 1 | 0 | 0 |
| $bp_i$ | 1 | 1 | 1 | 0 |
| $bx_i$ XOR $by_i$ | 0 | 1 | 0 | 0 |
| $bx_i$ AND $bp_i$ | 1 | 0 | 0 | 0 |
| Calculated $FIF_i$ | 1 | 1 | 0 | 0 |

Overall $FIF = FIF_0 \times FIF_1 \times FIF_2 \times FIF_3$
$= 1 \times 1 \times 0 \times 0 = 0$

State transition 2: INIT (1100) $\rightarrow$ LOAD1 (0000), protected_state (encoding=RESULT)

| $i$ | 0 | 1 | 2 | 3 |
|---|---|---|---|---|
| $bx_i$ | 1 | 1 | 0 | 0 |
| $by_i$ | 0 | 0 | 0 | 0 |
| $bp_i$ | 1 | 1 | 1 | 0 |
| $bx_i$ XOR $by_i$ | 1 | 1 | 0 | 0 |
| $bx_i$ AND $bp_i$ | 1 | 1 | 0 | 0 |
| Calculated $FIF_i$ | 1 | 1 | 0 | 0 |

Overall $FIF = FIF_0 \times FIF_1 \times FIF_2 \times FIF_3$
$= 1 \times 1 \times 0 \times 0 = 0$

... ... ...

</div>

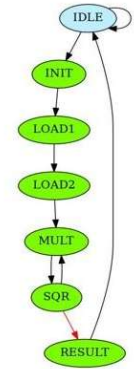

Fig. 6. State transition graph for the input design used in Case Study IV-A.

*2) Case Study IV-C: Security Assessment through Open-ended Question:* In the preceding investigations, namely Case Study IV-A, our emphasis was on providing detailed prompts to GPT-3.5, targeting security metric computations. This approach was necessitated due to the relative limitations of GPT-3.5 in executing advanced reasoning tasks. Contrarily, in the present study, we prefer conducting a security assessment on GPT-4 using open-ended questions, deliberately avoiding any predefined prompting strategies.

The FSM design under scrutiny in this study is presented in Listing 6. A thorough analysis of this design reveals that the state "s3" within the FSM stands isolated, devoid of any incoming transitions, making it an unreachable state. This configuration is not merely a design flaw; it poses a significant security risk. An attacker might leverage these states to communicate or transfer information without detection, thus compromising confidentiality. With this design as input, we prompted GPT-4, succinctly asking, "Is there any issue with this design?". GPT-4's feedback is visualized in Figure 7. Impressively, GPT-4 identified a multitude of design issues, including the critical observation of the unreachable state. Beyond this primary concern, GPT-4 further pinpointed other design elements needing improvement, showcasing its comprehensive design evaluation capabilities.

In our evaluations spanning various scenarios, including static deadlock, dynamic deadlock, absence of default statement, and race condition, we consistently employed open-ended questions with GPT-4. Remarkably, GPT-4 demonstrated a significantly higher success rate in precisely assessing security concerns, with the exception of dynamic deadlock situations. Through these case studies, it became evident that GPT-4 possesses the capability to deeply dissect and comprehend FSM designs, promptly identifying simple and not-so-complicated design imperfections. Furthermore, contrasting its predecessors, GPT-4 accentuates its advanced reasoning skills by adeptly responding to open-ended inquiries without the necessity of iterative prompting.

*E. Case Study VI: Countermeasure Development*

In this exploration, we aim to assess how adept GPT-4 is at identifying and fixing vulnerabilities in FSM designs. We provided GPT-4 with an FSM design that had clear breaches illustrative purposes, our chosen FSM design contains 7 states. The state transition graph (STG) of this design mirrors the one depicted in Figure 6. "Result" is considered as the protected state. The outcomes generated by *Prompt 3.1*, *Prompt 3.2*, and *Prompt 3.3* elucidate the step-wise systematic approach through which GPT-3.5 adeptly tackles the computationally intensive task of determining the FIF metric. To ensure clarity in our demonstration, we opted to abstain from displaying outputs for each individual state transition. A critical observation from this case study is the indispensable role of multi-stage prompting coupled with a structured tabular data representation in meticulously calculating such security metrics. This led to two more prompt guidelines for the calculation of extensive SoC security tasks.

*Prompt Guideline 5:* For security assessment tasks demanding complicated and comprehensive decision-making processes, adopt a multi-stage prompting methodology. In this approach, each prompt should target a concise, specific task, and the results from one task should seamlessly serve as the input for the subsequent task.

*Prompt Guideline 6:* When addressing tasks that involve extensive data computation, particularly during security metric calculations, structure the data using a tabular format for clarity and ease of processing.

```verilog
always @(*)begin
   next_state =s0;
   case (curr_state)
      s0: begin
         sbit =;
         if(start) next_state =s1;
         else next_state =s0;
      end
      s1: begin
         sbit =ctrl;
         if (ctrl) next_state =s4;
         else next_state =s2;
      end
      s2: begin
         next_state =s0;
      end
      s3: begin
         if(finish) next_state = s0;
         else next_state=s3;
      end)@*)
      s4: begin
         if (sbit) next_state =s1;
         else next_state =s2;
      end
   endcase
end
endmodule
```

Listing 6. Input design used in security assessment by GPT-4 through open-ended question

Fig. 7. Security assessment by GPT-4 through open-ended question

---

**Prompt 4.1**

<Input Design>

In this case, assume WAIT_KEY is the protected state and other states are unprotected.
There are two security rules:
1. All unused states of a control FSM should be handled through the 'default' statement in the RTL description
2. When state transition takes place between two consecutive unprotected states, the hamming distance between the states should be 1.

Security Assessment:
These two rules are violated in this design in the following way:
1. The is no 'defualt' statement through which unused states '101', '110', and '111' are handled.
2. There are following two state transitions between unprotected states where the hamming distance is not 1.
WAIT_DATA - INDIAL_ROUND : 001 - 010 : HD=2
DO_ROUND - FINAL_ROUND: 011 - 100: HD=3

Violation Mitigation Instructions:
Modify the FSM design so that the rules are followed. While modification the STG graph remains the same. For the modified design, check if is there any rule violation in the provided design. If yes, continue modifying until two rules are followed in the modified design

---

we refrained from providing GPT-4 with any specific mitigation strategies, requiring the model to derive solutions autonomously. A sample of the prompt is outlined in Prompt 4.1.

Listing 7 shows the input design used in this case study. Here, 'WAIT_KEY' is considered the protected state. From careful observation, it becomes evident that the input design has two violations of security rules. At first, there is no default statement to handle the unspecified states. This can lead to unpredictable behavior when the system encounters an unspecified state. This unpredictability can be exploited by attackers. Secondly, the hamming distance between unprotected states becomes greater than 1 in the following transitions: like 'WAIT_DATA' to 'INITIAL_ROUND' and 'DO_ROUND' to 'FINAL_ROUND'.

- WAIT_DATA - INITIAL_ROUND : Hamming Distance = 2
- DO_ROUND - FINAL_ROUND : Hamming Distance = 3
- FINAL_ROUND - WAIT_DATA : Hamming Distance = 2

of two important security rules related to state transitions and the use of default statements, as highlighted in [65]. Our approach in querying GPT-4 was strategic: we briefed the model about these security rules and the security assessment where we mentioned how the rules are violated in the provided design. Subsequently, we tasked GPT-4 with amending the design to ensure compliance with these rules. Additionally, we asked GPT-4 to review its own revised design, drawing from our approach in Case Study I. This method not only checks GPT-4's skill in addressing problems but also its ability to double-check its work, ensuring that corrections are both safe and aligned with recommended practices. Importantly,

```verilog
module fsm_module(KR, DS, clk, reset);
   input clk; // clock signal
   input reset; // reset input
   input KR, DS;
   parameter WAIT_KEY = 3'b000, WAIT_DATA = 3'b001,
      INITIAL_ROUND = 3'b010, DO_ROUND = 3'b011,
      FINAL_ROUND = 3'b100;
   reg [2:0] current_state, next_state;
   always @(current_state, KR, DS)
   begin
    case(current_state)
       WAIT_KEY: begin
          if(KR ==0) next_state =WAIT_DATA;
          else next_state =WAIT_KEY;
          end
       WAIT_DATA: begin
          if(DS ==0) next_state =WAIT_KEY;
          else next_state =INITIAL_ROUND;
          end
       INITIAL_ROUND: begin
          if(KR ==0) next_state =WAIT_KEY;
          else next_state =DO_ROUND;
          end
       DO_ROUND: begin
          if(KR ==0) next_state =WAIT_KEY;
          else next_state =FINAL_ROUND;
          end
       FINAL_ROUND:
          begin
          if(KR ==0) next_state =WAIT_KEY;
          else next_state =WAIT_DATA;
          end

    endcase
   end
   // sequential memory of the Moore FSM
   always @(posedge clk, posedge reset)
   begin
    if(reset ==0)
       current_state <=WAIT_KEY;
    else
       current_state <=next_state;
   end
endmodule
```

Listing 7. Input design used in Case Study VI

```verilog
module fsm_module(KR, DS, clk, reset);
   input clk; // clock signal
   input reset; // reset input
   input KR, DS;
   parameter WAIT_KEY = 3'b000, WAIT_DATA = 3'b001,
   INITIAL_ROUND =3'b011, DO_ROUND = 3'b010,
   FINAL_ROUND = 3'b110;
   reg [2:0] current_state, next_state;

   always @(current_state, KR, DS)
   begin
      case(current_state)
         WAIT_KEY: begin
             if(KR ==0) next_state =WAIT_DATA;
             else next_state =WAIT_KEY;
         end
         WAIT_DATA: begin
             if(DS ==0) next_state =WAIT_KEY;
             else next_state =INITIAL_ROUND;
         end
         INITIAL_ROUND: begin
             if(KR ==0) next_state =WAIT_KEY;
             else next_state =DO_ROUND;
         end
         DO_ROUND: begin
             if(KR ==0) next_state =WAIT_KEY;
             else next_state =FINAL_ROUND;
         end
         FINAL_ROUND:
             begin
             if(KR ==0) next_state =WAIT_KEY;
             else next_state =WAIT_DATA;
         end
             default: next_state = WAIT_KEY;
             // default to protected state
      endcase
   end

   // sequential memory of the Moore FSM
   always @(posedge clk, posedge reset)
   begin
      if(reset ==0)
          current_state <=WAIT_KEY;
      else
          current_state <=next_state;
   end
endmodule
```

Listing 8. Design generated by GPT-4 in Case Study VI

This scenario is particularly vulnerable to fault injection attacks. The generated design by GPT-4 is shown in Listing 8. It can be seen that the default statement has been successfully added to mitigate the security issue. For the second security breach, GPT-4 used gray encoding to restructure the current state encodings. This strategy successfully addressed two of the three highlighted vulnerabilities. However, the transition from FINAL_ROUND to WAIT_DATA remains a concern, with a Hamming distance of 3. While GPT-4 did not entirely eliminate this vulnerability, its approach did reduce the associated risk. The application of gray encoding, as suggested by GPT-4, is indeed a viable solution to rectify such design flaws. This demonstrates the ability of GPT-4 to autonomously derive effective solutions without external guidance. Ideally, to fully address the security concern, gray encoding should be applied to unprotected states initially, followed by the assignment of protected states. The solution provided by GPT-4 closely aligns with this optimal mitigation strategy. With more explicit context on the mitigation strategy, it is plausible that GPT-4 could have devised a complete countermeasure. It is worth noting that, the mitigation of the second security violation is complex. In some cases, it may be challenging to find an encoding arrangement that fully adheres to this security rule.

### F. Case Study VII: Testbench Generation

In the domain of security verification, the development of testbenches is often time-consuming, especially when dealing with substantial designs. Therefore, we present an investigation into the creation of test benches for security verification using ChatGPT. These automatically generated testbenches are aimed at detecting CWEs and other potential security vulnerabilities within a design. The principal objective of these generated testbenches is to identify counterexamples, which demonstrate violations of targeted CWEs or security rules.

To harness ChatGPT-generated test benches for security verification, it is crucial to validate their syntactic and semantic correctness, as well as their ability to detect security weaknesses. Consequently, we generated a set of examples to evaluate these testbenches. We pursued a comparative analysis by generating two versions of test cases: one incorporating the security weakness and the other devoid of it. Subsequently, we executed simulations of the design and the corresponding testbenches using Modelsim.

We first consider CWE-835, which highlights the security frailties associated with an unreachable loop. To expose this

```
module testbench;
    reg KR, DS, clock, reset;
    wire [2:0] current_state, next_state;

    FSM uut (.KR(KR), .DS(DS),
    .clock(clock), .reset(reset),
    .current_state(current_state),
    .next_state(next_state));

    integer i;

    initial begin
        clock =0;
        forever #5 clock =~clock;
    end

    initial begin
        reset =1;
        #10 reset =0;
    end

    initial begin
        // Wait for the FSM to reach the WAIT_KEY state
        while(current_state !=3'h1) begin
            #5;
        end

        // Generate input sequences to test for CWE 835
        KR =1; DS =0; // should transition to WAIT_KEY
        #5 KR =0;
        #5 KR =1; DS =1; // should transition to
        //INITIAL_ROUND
        #5 KR =0;
        #5 KR =1; DS =1; // should transition to DO_ROUND
        #5 KR =0;
        #5 KR =1; DS =1; // should transition to FINAL_ROUND
        #5 KR =0;
        #5 KR =1; DS =1; // should transition to WAIT_DATA
        #5 KR =1; DS =0; // should transition to WAIT_KEY

        // Check for CWE 835 by ensuring that the testbench
        //does not get stuck in a loop
        for(i =0; i <10; i =i +1) begin
            #5;
            if(current_state ==3'h1) begin
                $display("CWE 835 detected!");
                $finish;
            end
        end

        $display("No CWE 835 detected.");
        $finish;
    end

endmodule
```

Listing 9. CWE-835 Testbench Verilog code generate by ChatGPT.

security vulnerability, a test case generated within a testbench should illustrate entry into the loop without any possibility of exit. One example of the generated Verilog file of the Testbench by ChatGPT is demonstrated in Listing 9. In the context of CWE-835, the testbench successfully identified the presence of the unreachable loop and effectively flagged it within the printed message.

Turning to the CWE570 (expression is always false) scenario, where an always-false expression was examined, the testbench printed message shows that the final state was never reached but the conditions to reach that state were not satisfied. As it is shown in the figure 8, the reset is constantly active, making it impossible to transition to the next state. Therefore, the printed message that shows that the state was never reached is not a counterexample.

There are indeed several advantages to employing ChatGPT

for the automated generation of testbenches, particularly when compared to the manual authoring of these testbenches. LLM-generated testbenches offer notable advantages, especially in scenarios involving large and intricate designs. They are tailored specifically to a given design, enhancing efficiency. Furthermore, they exhibit versatility in language translation, with the ability to seamlessly transition between languages such as Verilog, SystemVerilog, and iVerilog, facilitated by straightforward prompts. Additionally, this approach accommodates design and verification engineers who may not possess expertise in a particular language, empowering them to craft testbenches in that language.

Nevertheless, our assessments have unveiled certain challenges associated with these automatically generated testbenches. Initial attention is required to ensure the syntactic and semantic correctness of the testbench. Subsequently, the design and associated testbench necessitate simulation in a separate software environment, such as Siemens ModelSim. Following this, a thorough analysis of the test output waveforms is essential to ascertain the successful detection or non-detection of vulnerabilities.

Our case studies revealed minor syntactic and semantic issues, including clock generation errors, design instantiation problems, and the instantiation of non-I/O port signals. Addressing these issues is feasible; however, a comprehensive assessment is necessary to gauge the test bench's effectiveness in confirming the presence of security vulnerabilities. Although ChatGPT-generated test benches showed promising results, their overall capacity to detect security vulnerabilities remains inadequate. This observation parallels the conclusions drawn from a previous study on ChatGPT-generated test benches for verification of functionality in ChipGPT [72]; it revealed challenges in their creation and the need for modifications. These limitations are rooted in the insufficient availability of suitable training data within the test bench and verification code domain.

***Prospect 10:*** Through tailored prompts, targeting specific security weaknesses within test cases, the automated generation of testbenches using LLM technology displays substantial potential to notably improve the speed, efficiency, and adaptability of testbench creation dedicated to security vulnerability detection.

***Challenge 14:*** Generating testbenches with LLMs is fast and easy, but assessing their correctness and the accuracy of vulnerability detection remains a challenge. Developing an automated workflow for design simulation and vulnerability detection is promising but demands significant effort.

## G. Case Study VIII: Security Assertion

The importance of security properties in formal verification is of utmost importance as they play a critical role in guaranteeing the strength and dependability of complex systems. In this case study, our objective is to demonstrate the ability of GPT-3.5 to understand security properties, expressed in natural language and presented as SVAs. To accomplish this, we presented GPT-3.5 with a collection of security properties formulated in SVA and requested it to provide a general

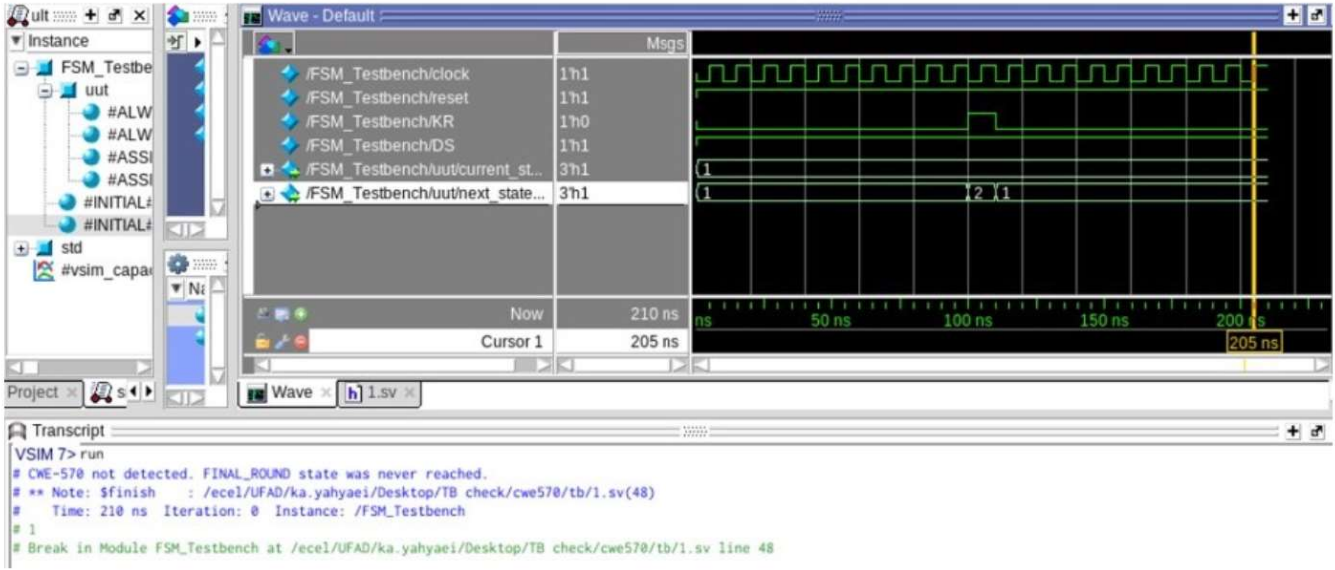

Fig. 8. Failed testcase of testbench generation for security verification

**Prompt 5.1**

$property\ p1;$
$@(posedge\ clk)RSA\_binary.FSM ==$
$RSA\_binary.IDLE\ \&\&\ rose(start)$
$|=> (RSA\_binary.FSM == RSA\_binary.INIT)$
$\#\#1(RSA\_binary.FSM == RSA\_binary.LOAD1)$
$\#\#1(RSA\_binary.FSM == RSA\_binary.LOAD2)$
$\#\#1(RSA\_binary.FSM == RSA\_binary.MULT)$
$\#\#1(RSA\_binary.FSM == RSA\_binary.SQR)$
$\#\#1(RSA\_binary.FSM == RSA\_binary.MULT)$
$\#\#1(RSA\_binary.FSM == RSA\_binary.SQR)$
$\#\#1(RSA\_binary.FSM == RSA\_binary.MULT)$
$\#\#1(RSA\_binary.FSM == RSA\_binary.SQR)$
$\#\#1(RSA\_binary.FSM ==$
$RSA\_binary.RESULT);$
$endproperty$

**Response to Prompt 5.1**

In conclusion, this property verifies the sequence of states that the FSM should allow upon the rising edge of the `start` signal. This can be useful to ensure that the FSM is progressing through its states as expected, and it can help detect any issues with the state transitions.

extract and convey the underlying intent of these assertions. As an example, Prompt 5.1 showcases one of the SVA-based security properties given to GPT-3.5 for a design of the controller of RSA shown in Figure 9. GPT-3.5 was asked what this property actually checks. "Response to Prompt 5.1" displays the final portion of the reply by GPT-3.5. From the response, it becomes evident that GPT-3.5 not only recognizes the structural components of the property but also grasps the semantic implications of the assertion. The response of the model accurately captures the essence of the property, highlighting its role in verifying the correct sequence of state transitions in the FSM. This shows the potential of GPT-3.5 as a tool for understanding and possibly helping design and verification of complex digital systems.

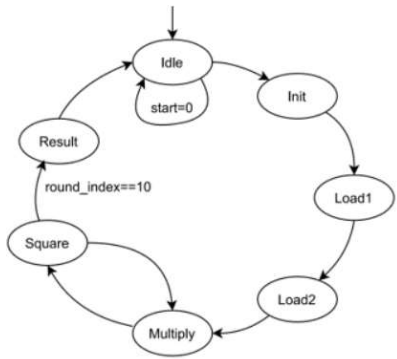

Fig. 9. Controller of RSA used in Case Study VIII

grasp of these assertions. Our goal was to discern whether the model could not only parse the technical syntax but also

## VI. LARGE-SCALE INVESTIGATION ON GPT CAPABILITIES

In Section V, we presented seven different case studies illustrating the potential of GPT-3.5 and GPT-4 to perform various SoC security tasks. However, these case studies alone are not sufficient to draw definitive conclusions. To address the research questions posed in Section IV, a more extensive examination of GPT's capabilities is essential. Consequently, in this section, we undertake a comprehensive investigation into the abilities of GPTs, focusing on four key areas: insertion of security vulnerability, detection of security rule violation, identification of hardware threat, detection of coding issues, and development of countermeasures.

Before discussing these security tasks, it is important to assess the proficiency of GPT-3.5 in generating syntactically correct hardware designs. In our analysis, we examined the 1806 Verilog designs produced by GPT-3.5. A meticulous syntax check revealed that 1580 of these designs were devoid of syntax errors. The remaining designs exhibited minor syntactical oversights, such as extraneous punctuations, omitted parentheses, usage of obsolete constructs, and missing 'begin' blocks. Additionally, there were instances where there was a mix-up between Verilog and SystemVerilog constructs. It should be noted that the syntax issues identified in the designs are relatively minor and can be rectified with ease. Such errors are often common even among experienced designers, especially during initial design drafts. Automated tools or linting software can quickly identify and correct these issues, ensuring the final design is both syntactically and semantically correct. Even GPTs themselves can do this correction by self-scrutinizing the design as demonstrated in Case Study I.

*Prospect 11:* The minor syntactical oversights present in the GPT-generated designs highlight an opportunity for integrating automated tools or linting software, or even enhancing GPT models for self-scrutiny, to ensure the generation of error-free, optimized hardware designs

### A. Insertion of Vulnerability

A database containing vulnerable designs is crucial in the hardware security domain. Such a database can serve as a rich resource for training and testing security tools, enabling them to better recognize and address vulnerabilities in hardware design. However, there is a scarcity of such databases in the community because the insertion of vulnerability in hardware design is a complicated task. It requires advanced knowledge, effort, and time to create such a database manually. That motivates us to investigate the proficiency of GPT-3.5 in embedding various vulnerabilities and CWEs into hardware designs.

In this context, we selected five distinct security vulnerabilities, each with its own security implications. These vulnerabilities are as follows:

- CWE 835 (Unreachable Exit Condition): This vulnerability can lead to infinite loops or prolonged execution, potentially causing a denial of service, resource exhaustion, and operational disruptions.
- Unused states NOT handled through the 'default' statement: This can result in unpredictable system behavior when an undefined state is encountered, potentially leading to system crashes or unintended operations.
- Duplicate Encoding State Integrity: This vulnerability can cause ambiguity in state interpretation, leading to unintended transitions or actions and compromising the system's reliability.
- Presence of Unreachable State Integrity: The existence of states that cannot be reached in any operational scenario might indicate design inefficiencies or latent bugs. Such states can lead to wasted computational resources, increased power consumption, or even mask other undetected vulnerabilities due to these states' inaccessibility.
- Presence of static deadlock: This can halt system operations, causing a standstill in processes and potentially leading to system timeouts or failures.

For vulnerability insertion, in each case, a specific prompting technique has been applied. Table VI shows the performance of GPT-3.5 in the successful insertion of these vulnerabilities. With the exception of the unreachable state scenario, GPT-3.5 boasts a success rate exceeding 85% in integrating these security flaws into the designs. The most notable success has been observed in scenarios involving the 'default' statement, while the insertion of unreachable states into FSM designs posed the greatest challenge. This can be attributed to the nuanced definition of an unreachable state in an FSM, described as a state devoid of any input transition condition but possessing one or more transitions to other states [65]. This definition bears a striking resemblance to the definitions of both dead state and static deadlock, leading to potential ambiguities. Such ambiguities, in turn, inadvertently introduce other security vulnerabilities during this task, leading to a lower success rate. It is noteworthy that GPT-3.5 attains 88.27% success in creating complicated situations like static deadlock. It involves the same strategy that we applied in Case Study II. It clearly indicates that through proper guidance, LLMs can be used to insert complicated vulnerabilities and weaknesses into hardware designs.

*Prospect 12:* The superior performance of GPT-3.5 in the successful insertion of hardware vulnerabilities and weaknesses into hardware design suggests a promising avenue for utilizing LLMs in creating databases of vulnerable designs, essential for developing security tools

### B. Detection of Security Rule Violation

In order to assess the performance of vulnerability detection, we selected the following three security rule violations in FSM designs defined by [65]

- A state with a static deadlock scenario must not exist.
- Each state must be encoded uniquely.
- When state transition occurs between two consecutive unprotected states, the Hamming Distance (HD) between those should be '1.'

In this task, the input design might either adhere to or violate certain security rules. The primary objective was to determine whether a given rule had been violated. As detailed in Case Study IV, we designed specific prompting strategies to detect any security rule violations. Analogously, distinct prompting methodologies were employed for each rule in this context. To estimate the accuracy of decisions made by GPT-3.5, we cross-referenced its decisions with the established ground truth.

Table VII presents a comparative analysis of the performance of GPT-3.5 and ARC-FSM in detecting embedded vulnerabilities within design inputs. For each violation, the table lists the number of input designs tested, the count of accurately detected vulnerabilities, and the corresponding accuracy percentage for both frameworks. From the table, it is evident that while GPT-3.5 demonstrates commendable accuracy in detecting vulnerabilities, ARC-FSM consistently exhibits higher or near-perfect accuracy rates across the tested security rules. For instance, in the case of "Presence of static deadlock," GPT-3.5 achieved an accuracy of 79.12%, whereas ARC-FSM reached 88.64%. Similar trends are observed for other security rule violations, underscoring the effectiveness of specialized tools like ARC-FSM in vulnerability detection. However, it is noteworthy that GPT-3.5 still offers competitive performance, especially considering its general-purpose nature compared to the specialized design of ARC-FSM.

TABLE VI
Performance of GPT-3.5 in successfully insertion of different hardware vulnerabilities and weaknesses.

| Vulnerability | CIA Violation | # Input Designs | # Successful Insertion | Success Rate (%) |
|---|---|---|---|---|
| CWE 835 (Unreachable Exit Condition) | Availability | 166 | 154 | 92.77 |
| All unused states NOT handled through the 'default' statement | Availability | 152 | 143 | 94.08 |
| Duplicate Encoding State | Integrity, Availability | 251 | 220 | 87.64 |
| Present of unreachable state | Integrity Availability | 251 | 161 | 64.14 |
| Present of static deadlock | | 273 | 241 | 88.27 |

TABLE VII
Comparison of performance of GPT-3.5 and ARC-FSM in the detection of vulnerabilities embedded into designs.

| Security Rule Violation | Framework | # Input Designs | # Accurate Detection | Accuracy (%) |
|---|---|---|---|---|
| Presence of static deadlock | GPT-3.5 | 273 | 216 | 79.12 |
| | ARC-FSM | | 242 | 88.64 |
| Duplicate Encoding State | GPT-3.5 | 351 | 322 | 91.74 |
| | ARC-FSM | | 321 | 99.69 |
| HD between unprotected state transitions | GPT-3.5 | 209 | 172 | 82.30 |
| | ARC-FSM | | 209 | 100.00 |

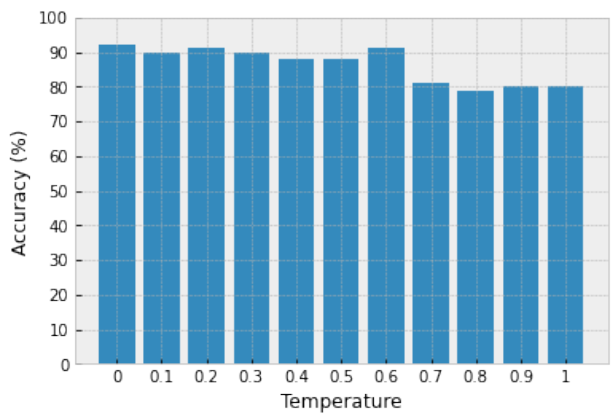

Fig. 10. Impact of 'temperature' parameter of GPT-3.5 in the detection of "duplicate state encoding".

In all of the above-mentioned experiments, we set the temperature of GPT-3.5 very low to keep the model deterministic. But intuitively temperature should have an impact on the performance of vulnerability detection. As we mentioned before, the 'temperature' parameter in language models like GPT-3.5 essentially controls the randomness of the model's predictions. A lower temperature steers the model towards more deterministic outputs, while a higher value encourages diversity, potentially leading to more creative but less precise responses. In order to investigate this impact, we vary the temperature of GPT-3 from 0 to 1 with an increase of 0.1 for the task of detecting the violates of 'unique encoding state' in 100 input designs. The impact of the 'temperature' parameter is shown in Figure 10. This suggests that a higher temperature lowers the accuracy of the model in this specific context. However, it would be premature to conclude that a lower temperature is universally ideal for vulnerability detection tasks. We are not advocating for the exclusive use of a lower temperature setting; instead, we emphasize that The impact of temperature is multifaceted and warrants meticulous investigation to harness the optimum performance of LLMs. The observation emphasizes the importance of carefully tuning the temperature parameter, especially in critical tasks such as vulnerability detection.

Furthermore, it is worth noting the inherent variability in the performance of GPT. In repeated testing iterations, GPT does not consistently yield identical performance metrics. This variability emphasizes the non-deterministic nature of the LLM and the challenges it poses. However, our experiments also revealed a silver lining: by employing detailed and stringent prompting strategies, the consistency and stability of the results can be significantly enhanced. Although GPT models offer immense potential, their application in critical domains such as vulnerability detection requires a nuanced approach. A thorough understanding of parameters such as temperature, combined with rigorous prompting strategies, is essential to realize their full potential while ensuring reliability.

***Achievement 8:*** GPT-3.5 has demonstrated commendable accuracy in detecting security rule violations in FSM designs, showcasing its applicability even when compared to specialized tools

***Challenge 15:*** The non-deterministic nature of LLMs, highlighted by variability across testing iterations, poses a challenge in achieving consistent results.

TABLE VIII
PERFORMANCE OF PROPOSED GPT-3.5 AND GPT-4 IN
HARDWARE TROJAN DETECTION

| LLM | Test Method | Total # Tests | # Detected Trojans | Detection Accuracy (%) |
|---|---|---|---|---|
| GPT-3.5 | Blind Test | 275 | 0 | 0 |
| | Contextual Test | 275 | 45 | 16.36 |
| GPT-4 | Blind Test | 216 | 175 | 81.01 |
| | Contextual Test | 108 | 99 | 91.67 |

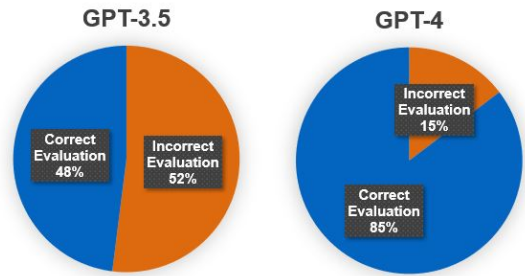

Fig. 11. Experimental results of GPT-3.5 and GPT-4 in identifying simple coding issues in 75 different hardware designs.

*Prospect 13:* There is potential to enhance the consistency and accuracy of GPT models in detecting vulnerabilities by meticulously tuning parameters like temperature and employing rigorous prompting strategies.

### C. Hardware Trojan Detection

Hardware Trojans present grave risks to electronic systems, with potential consequences ranging from unauthorized access and data breaches to complete system malfunctions. Detecting these Trojans is imperative, especially when considering the security of national critical infrastructures or defense systems. In our study, we investigated the abilities of GPT-3.5 and GPT-4 to identify hardware Trojans within AES designs, utilizing 28 distinct AES designs sourced from the Trust-Hub Trojan Benchmark [228].

To enhance the rigor of our investigation, we first sanitized the RTL code of any overt indications of Trojan presence. This involved renaming Trojan modules and eliminating explicit terms like "Trojan" and "trigger" from the RTL. Our assessment employed two distinct methodologies: the Blind Test and the Contextual Test. The former tests the raw capabilities of the models by providing no explicit context about the design. Conversely, the Contextual Test offers models supplementary information, outlining potential Trojan insertion techniques in AES designs. To further the depth of our analysis, each test setup was executed under varying temperature settings, incrementally adjusted from 0 to 1 at intervals of 0.1. This process ensures the thoroughness of the investigations.

Table VIII provides a comparative analysis of the hardware Trojan detection capabilities of GPT-3.5 and GPT-4. From the result, it is very clear that GPT-3.5 does not have enough knowledge to detect such a security threat. When enhanced knowledge is given to it, it attains 16.36% accuracy which is still low. The length of the context also becomes a problem for GPT-3.5. Due to the comparatively shorter context length of the model, it becomes challenging to analyze the whole AES design with enough context with GPT-3.5. On the other hand, GPT-4 significantly outperforms GPT-3.5 with an impressive detection accuracy of 81.01%, even when no context is provided. The contextual test, which provides the models with additional knowledge, further accentuates the superiority of GPT-4, achieving a 91.67% accuracy rate. These results clearly indicate the advances in GPT-4, highlighting its improved proficiency in hardware Trojan detection over its predecessor. Such superior performance of GPT-4 in both tests indicates a deeper understanding of hardware designs and threats, emphasizing its potential as a valuable tool in hardware security evaluations.

The performance of GPT-4 in the detection of hardware Trojans is on par with other machine learning-based algorithms [229]. Traditional ML approaches necessitate a series of steps including data pre-processing, feature extraction, and model training. On the contrary, LLMs like GPT-4 streamline this process, achieving comparable results through simple natural language descriptions. The efficacy of LLMs can be further enhanced with refined prompting or task-specific fine-tuning. A distinct advantage that GPT-4 offers in this task is its ability to address design dependencies, a prevalent challenge in conventional ML-based hardware Trojan detection methods. For example, in the case of ML-based approaches, a model trained on a dataset of Trojan-injected AES designs may not effectively detect Trojans within other IP core designs. LLMs like GPT-4 have the capability to solve this issue. Because GPT-4 is equipped with advanced natural language processing capabilities that enable it to understand and analyze the contextual and structural intricacies inherent in hardware designs. It can process and interpret the natural language descriptions of hardware designs, allowing for a more holistic and nuanced analysis. This capability facilitates a more comprehensive detection of potential Trojans.

*Achievement 9:* GPT-4 has achieved hardware Trojan detection performance comparable to traditional ML algorithms, but with a simplified process utilizing natural language descriptions.

### D. Detection of Coding Issue

In the domain of hardware design, even seemingly minor coding issues can have profound implications. While these issues, such as linting discrepancies, structural anomalies, coding style inconsistencies, and synthesis problems, might not directly manifest as security vulnerabilities or weaknesses, they can act as precursors. Such coding issues can inadvertently introduce or propagate more severe security vulnerabilities in the design. Ensuring the absence of these coding issues is not just about maintaining a clean codebase; it is also about preempting potential security risks and ensuring the robustness and reliability of the hardware. For this experiment, we sourced 75 coding issues and the test designs from the Jasper Superlint reference manual [213]. It is crucial to note that we took meticulous measures to sanitize the designs. We

| Vulnerability | CIA Violation | # Input Designs | # Successful Mitigation | Success Rate (%) |
|---|---|---|---|---|
| Duplicate Encoding State | Integrity, Availability | 161 | 147 | 91.30 |
| Present of unreachable state | Integrity Availability | 160 | 158 | 96.43 |
| Present of static deadlock | Availability | 168 | 162 | 96.43 |

ensured these designs lacked explicit indications or markers pointing toward the coding issue.

Figure 11 offers a comparative analysis of GPT-3.5 and GPT-4 in pinpointing these coding issues within hardware designs. Of the 75 different hardware designs evaluated, GPT-3.5 accurately detected coding issues in nearly 48% of the cases. While this indicates a moderate capability of GPT-3.5 in identifying such issues, there is evident potential for enhancement. In contrast, GPT-4 showed proficiency in accurately detecting issues in 85% of the designs, a significant improvement 37% over GPT-3.5. This marked advancement suggests that GPT-4 has a more remarkable ability to discern and highlight coding problems in hardware designs.

*Achievement 10:* GPT-4 has shown a significant improvement in identifying and addressing coding issues, marking notable progress in the field of automated code analysis.

### E. Vulnerability Mitigation

Vulnerabilities within hardware designs can be particularly detrimental, given their foundational role in many electronic systems. Recognizing this, we have examined the capabilities of GPT-3.5 to rectify these security vulnerabilities in FSM designs. In our experimental setup, we took a directed approach to harness the capabilities of GPT. Instead of leaving the model to blindly identify vulnerabilities, we explicitly informed GPT about the specific security vulnerability present within the design. With this knowledge, GPT was then tasked with the challenge of mitigating the identified vulnerability. In this experiment, we have considered three distinct vulnerabilities: duplicate encoding state, presence of unreachable state, and presence of Static Deadlock. Each vulnerability is associated with potential violations of the CIA (Confidentiality, Integrity, and Availability) triad, underscoring its criticality in the context of secure hardware design. Table IX shows the performance of GPT-3.5 in mitigating these vulnerabilities.

For the problem of duplicate encoding state, remarkably, successful mitigation was achieved in 147 of 161 cases, translating to a success rate of 91.30%. This suggests that the mitigation techniques employed are highly effective against this particular vulnerability, ensuring that states in the design are uniquely encoded and, thereby, preventing ambiguous state transitions. The mitigation technique showed even greater efficacy in the case of the unreachable state, with 158 designs successfully rectified, resulting in a success rate of 96 43%. Unreachable states can indicate design inefficiencies or hidden vulnerabilities, so their effective mitigation is paramount to the overall robustness of the system. Lastly, the presence of static deadlock, which primarily affects availability, was evaluated in 168 designs. Deadlocks can halt system operations, leading to potential system failures. The mitigation techniques employed demonstrated a consistent success rate of 96.43%, with 162 designs successfully addressed. The findings highlight the robustness and effectiveness of the employed mitigation techniques against common hardware vulnerabilities. The consistently high success rates across different vulnerabilities emphasize the scope of employing LLM in vulnerability mitigation to ensure secure and reliable hardware designs.

*Prospect 14:* The high success rates in mitigating vulnerabilities open the door to further exploration and utilization of LLMs like GPT-3.5 in automated vulnerability mitigation, potentially revolutionizing the field of hardware security.

## VII. CONCLUDING REMARKS

The rapid advancements in the SoC domain and their pervasive presence in modern electronics systems have accentuated the urgency for robust and innovative security solutions. As SoCs become integral to many devices, from smartphones to autonomous vehicles, their security challenges become increasingly multifaceted. In parallel, the emergence and evolution of LLMs have revolutionized the field of NLP and even coding and reasoning tasks. With their unparalleled linguistic and reasoning capabilities, these models offer a promising avenue for addressing the sophisticated challenges of SoC security. Recognizing this potential, our research is motivated to dive deep into the confluence of LLMs and SoC security, aiming to harness the strengths of these models to perform SoC security tasks.

Throughout this work, we embarked on a comprehensive exploration of the role LLMs can play in various SoC security tasks. Our extensive survey of existing LLMs provided a detailed panorama of their development trajectories, capabilities, and potential applications. We have established vital research questions in this domain. To address these questions, we have demonstrated practical case studies and showcased different LLM capabilities in scenarios such as vulnerability insertion, security assessment, security verification, and countermeasure development. These case studies were rigorous evaluations of large-scale open-source designs that indicated both the potential and challenges of integrating LLMs into the SoC security framework. We also offered strategic guidelines for prompt engineering and presented large-scale investigations as a roadmap for future endeavors in this domain. Throughout this work, we identified 10 achievements, 15 challenges, and 14 prospects of LLM in SoC security validation. By bridging the gap between LLM capabilities and SoC security needs, we have laid a robust foundation that researchers and industry professionals can build upon.

The future of LLMs in SoC security looks promising as we look ahead. These advanced systems keep improving and more flexible, opening up many possibilities in SoC security. Although our research has made progress, there is still a lot that we need to explore. We hope our work guides and inspires more research, encourages new ideas, and helps collaborations in this growing field.